# Pulsations along the Asymptotic Giant Branch (AGB): a global view

*Pierre Darriulat, Pham Thi Tuyet Nhung and Do Thi Hoai*
*CASE, Viet Nam National Space Center, Vietnam Academy of Science and Technology,*
*18 Hoang Quoc Viet, Nghia Do, Ha Noi, Viet Nam. Email: darriulat@vnsc.org.vn*

**Abstract.** In three preceding articles, we studied the relation between the light curves of Mira variables and their stellar properties. The analysis was restricted to a sample of stars offering high density and quality of observations in the visible, allowing for the definition of simple parameterizations of the light curves. Clear relations were revealed between the evolution of the stars along the AGB and parameters such as the width of the light pulse or the deviation of its ascending branch from a sine wave. However, the restriction to Mira variables was found to be a major drawback preventing a complete picture to be drawn. The present article overcomes this deficiency by reviewing a large sample of semi-regular and irregular variables. It gives evidence for two families of stars having different progenies. The first family enters the AGB already pulsating with amplitudes at the scale of one unit of magnitude. It keeps cooling down and expanding, while pulsating with increasing amplitudes and periods and decreasing width of the light pulse; its progeny is dominated by some of the Mira variables studied in the three preceding articles. The second family enters the AGB before displaying significant pulsations; some of its progeny may leave the AGB without entering the Thermally Pulsing Branch (TPAGB) and experiencing strong Third-Dredge-Up (TDU) events; others may instead become Mira variables studied in the three preceding articles and pulsating with a broader light pulse. The possible transition of both families from oxygen-rich to carbon-rich is studied and described. Comments on the interpretation of these results in the context of current state-of-the-art knowledge are presented. While the stars of the first family tend to have larger initial masses than the stars of the second family, details of the properties of the convective envelope seem also to play a role in defining which of the two families a star belongs to.

## 1. Introduction

Stars ascending the Asymptotic Giant Branch (AGB) are the object of numerous studies addressing a broad range of questions. Some studies aim at understanding the mechanisms governing mass-loss, including effects such as of shock waves, dust condensation and related radiation pressure, impact of possible stellar or planetary companions, etc... Other studies aim at understanding the physics governing the interior of the star; in contrast with the former, which are based on observations, the latter must make use of models. In particular, the study of stellar pulsations aims at describing radial movements in the interior of the star, but has only access to information related to what happens on or above the photosphere: light curves measured at various wavelengths and radial velocities obtained from Doppler shifts. The present article explores how much can be learned from the observation of the light curve about the basic parameters that characterise the state of a star. As such a topic touches on several domains of stellar physics, we recall basic knowledge of relevance in the present introduction: on stellar evolution in Subsection 1.1 and on stellar pulsations in Subsection 1.2.

### *1.1 Stellar evolution*

Stars that ascend the AGB are approaching the end of their lives. They go through different successive phases that have increasingly shorter durations. It is therefore convenient to describe their evolution as a function of time counted backward from death, as done in Figure 1. The state of a star is essentially dependent on three parameters: its initial mass and metallicity and its current age. The dependence on metallicity is much smaller than that on the other two parameters and when dealing with stars in the solar neighbourhood, one can usually assume that they have metallicity close to solar ($Z$~0.015). We shall therefore focus on such stars in what follows. We define $M_i$ as the ratio of the initial mass (zero age main sequence) to the solar mass. We summarize below the main steps of the evolution of stars, starting from the less massive. Relevant references can be found in reviews such as Herwig (2005) and Karakas & Lattanzio (2014).

The duration of the protostar phase in Figure 1 is always negligible, less than 1% of the star lifetime. Red dwarfs, with 0.08< $M_i$ <0.7, fuse hydrogen to helium on the MS, on which they stay for a time exceeding the current age of the Universe. They make up three-quarters of the fusing stars in the Milky Way. The more massive red dwarfs, having $M_i$ >~0.23, may expand into red giants, but will never reach the tip of the red-giant branch and ultimately will become white dwarfs.

Stars having $M_i$ between ~0.7 and ~8 ascend the AGB. They stay on the MS until having exhausted hydrogen in the core, which becomes therefore a helium core. On the MS, stars having $M_i$ <~1.5 have convective envelopes, radiative cores and are primarily powered by the pp cycle. Conversely, stars with $M_i$ >~1.5 have radiative envelopes, convective cores and are primarily powered by the CNO cycle. All stars having $M_i$ between ~0.7 and ~8 leave the MS by starting to burn hydrogen in a shell around the helium core. For $M_i$<~1.5 the helium core becomes degenerate before fusing, while for $M_i$>~1.5 it fuses before becoming degenerate (we recall that degenerate means that the electrons in the core are no longer attached to specific nuclei but form a gas at the scale of the whole core, implying a strong increase of the density). But in both cases the star expands and cools and the core increases in mass as the shell produces more helium. This is the Red Giant Branch (RGB), on which the helium core grows until it contracts rapidly when helium starts to fuse in it, marking the transition to the Helium Core Burning (HeCB) phase. The time spent on the RGB becomes very short (negligible on Figure 1) when $M_i$ exceeds ~1.5 to 2, in which case the star transits promptly from MS to HeCB causing a gap in the Hertzsprung-Russel diagram called the Hertzsprung Gap (HG). In practice, we may distinguish between stars having low and high values of $M_i$, the former experiencing a significant RGB episode and the latter transiting smoothly from MS to HeCB; the border between the two regions depends on metallicity and on which particular process is being considered; for simplicity we refer to it as $M_i$~1.5 in what follows.

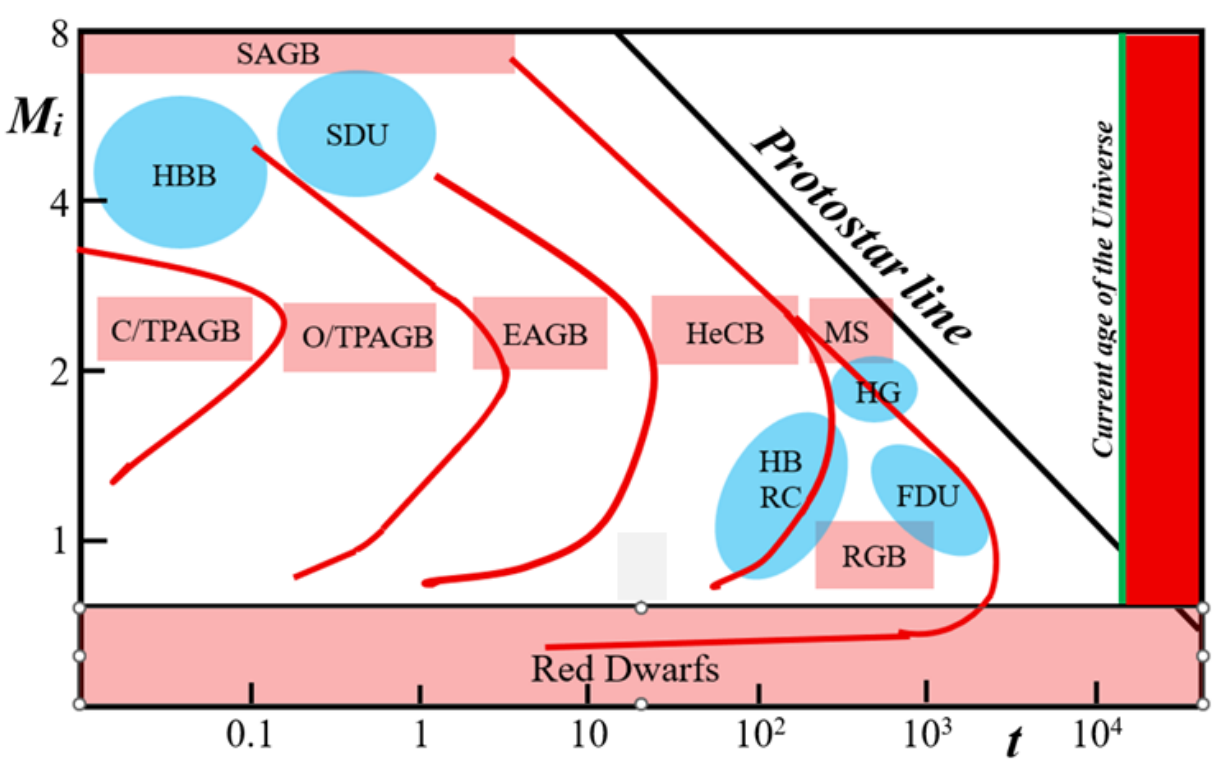

**Figure 1.** Sketch of stellar evolution in the initial mass $M_i$ (in units of solar masses) vs age *t* (in Myr counted backward from death time) plane, using Table 2 of Miller Bartolomi (2016) for a star of near-solar metallicity, *Z*=0.02. Death time is defined as the time when the mass of the envelope has decreased to 1% of the mass of the star. The main steps are shown as MS (Main Sequence), RGB (Red Giant Branch), HeCB (Helium Core Burning), EAGB (Early Asymptotic Giant Branch) and TPAGB (Thermally Pulsing Asymptotic Giant Branch), the latter being split between oxygen-rich (O/TPAGB) and carbon-rich (C/TPAGB). Specific episodes described in the text include HG (Hertzsprung Gap), FDU (First Dredge-Up), HB (Horizontal Branch), RC (Red Clump), SDU (Second Dredge-Up), HBB (Hot Bottom Burning) and SAGB (Super-AGB). The red arc between RGB and HeCB shows the Helium Core Flash. The black line is the Zero Age Main Sequence, covering the whole protostar phase and measuring the star lifetime.

For stars having $M_i$<~1.5, the transition from MS to RGB is followed by the occurrence of the First Dredge Up (FDU): fusion of hydrogen into helium now occurs in a shell surrounding the contracting degenerate helium core and the outer layers expand and cool, becoming convective due to the increase of opacity at lower temperatures, leading to mixing of the outer envelope with regions that have experienced some nuclear processing. The transition from RGB to HeCB for stars having $M_i$<~1.5 is marked by the Helium Core Flash when the core transits from degenerate to fusing He into carbon via the three-alpha cycle, while hydrogen keeps fusing into helium via the CNO cycle in a shell around the core. It is a particularly violent and brief flash because it occurs in a degenerate environment in contrast with the helium flashes (thermal pulses) of the TPAGB. The star spends then time at nearly constant luminosity either on what is called the Horizontal Branch (HB) if the metallicity is low or the Red Clump (RC) if the metallicity is closer to solar.

In the HeCB phase, stars having $M_i$ between ~0.7 and ~8 burn helium into carbon in the core via the three-alpha cycle, progressively forming a convective carbon and oxygen core: helium burning increases the content of $^{12}$C, which in turn increases the abundance of $^{16}$O from the reaction $^{12}$C(α,γ)$^{16}$O. When the helium supply in the core gets exhausted, the star enters the early AGB (EAGB) phase, with hydrogen and helium burning in concentric shells around what has become a C/O core. At that stage, for $M_i$>~4, the convective envelope grows into the stellar interior, producing the Second Dredge-Up (SDU); in contrast to the FDU, the SDU goes deeper and mixes material exposed to complete hydrogen burning.

During the EAGB phase, the main source of energy is helium fusion in a shell around a core consisting mostly of carbon and oxygen. When the helium shell runs out of fuel, the star enters the Thermally Pulsing AGB (TPAGB) phase, during which the star fuses hydrogen in a thin shell around the helium shell, first very thin but progressively enriched by the ashes of the hydrogen-burning shell surrounding it. Over periods of 10 to 100 kyr, the helium shell ignites explosively, producing a helium shell flash (TP, Thermal Pulse) causing material from the core region to be eventually mixed into the outer layers, in a process referred to as Third Dredge-Up (TDU). Sufficiently strong TDU events will bring s-process elements to the star surface and increase the carbon to oxygen ratio of the outer envelope, possibly transforming it from oxygen-rich to carbon-rich. In Figure 1 (*Z*=0.02) the number of thermal pulses during the TPAGB phase increases from 4 for $M_i$=1 to 34 for $M_i$=4. However, stars having $M_i$>4 experience Hot Bottom Burning (HBB): proton-capture nucleosynthesis at the base of the outer envelope favours the conversion of C to N by the CN-cycle and reconversion of the C-rich to an O-rich atmosphere. HBB is activated when the temperature at the bottom of the envelope reaches 40 MK, the same value as required to produce lithium, the presence of which offers therefore an efficient tool to help with the identification of HBB stars.

When on the TPAGB, the star develops stronger winds accelerated by radiation pressure on grains of dust that could form in the cooler layers of the atmosphere, causing the mass-loss rate to increase and the remaining lifetime to shorten. The structure of an AGB star is qualitatively the same for all masses. In summary, the TPAGB includes four phases: 1) Thermal pulse, the He shell burning brightly with up to ~$10^8$ $L_{sun}$ for ~100 years and driving a convective zone in the He intershell; 2) the power-down phase when the He shell dies down, the thermal pulse energy driving the expansion of the whole star and the extinction of the H shell; 3) the TDU, mixing, when strong enough, carbon and other He-burning products to the stellar surface; 4) the interpulse phase (~$10^4$-$10^5$ years) in between thermal pulses where the H shell provides most of the surface luminosity.

Stars with $M_i$>7 are called super-AGB stars and represent a transition to the more massive supergiant stars that undergo full fusion of elements heavier than helium. During the triple-alpha process, some elements heavier than carbon are also produced: mostly oxygen, but also some magnesium, neon, and even heavier elements. Super-AGB stars develop partially degenerate carbon–oxygen cores that are large enough to ignite carbon in a flash analogous to the earlier helium flash. The second dredge-up is very strong in this mass range and the size of the thermal pulses and third dredge-ups are reduced compared to lower-mass stars, while the frequency of the thermal pulses increases dramatically. Most Super-AGB stars end their life as oxygen–neon white dwarfs.

Finally, stars with $M_i$>8 quickly and smoothly initiate helium-core fusion after having exhausted their hydrogen in the MS and continue fusing heavier elements after helium exhaustion until they develop an iron core, at which point the core collapses to produce a supernova. After having left the MS, these stars are referred to as Super-Giants. They may pulsate with periods of a few years and amplitudes of a few units of magnitude (Heger 1997), as VX Sgr and S Per. Supernova explosions produce a neutron star or a black hole. The largest stars of the current generation have $M_i$~100 to 150.

Unfortunately, none of the basic parameters governing the star evolution (age, initial mass and metallicity) is directly accessible to observation. What is observed is essentially the spectrum of the light emitted by the photosphere and atmosphere of the star and their radial velocities. When the distance from us to the star is known, one can measure the absolute magnitude of its luminosity; to the extent that one has control over absorption, one can measure its dependence on wavelength and obtain an evaluation of the effective temperature. Observation therefore provides the location of the star in the luminosity vs temperature plane, referred to as the Hertzsprung-Russell (HR) diagram. What is a simple straight line parallel to the time axis in the $M_i$ vs $t$ plot of Figure 1 has now become a tortuous line in the HR diagram, illustrating the outstanding complexity and difficulty of obtaining estimates of the basic evolution parameters from observation. Additional information that helps with this task includes detection of different spectral lines in the spectrum, high spatial resolution observations of the photosphere and atmosphere using modern instruments such as ALMA and VLT, and, in the case of pulsating stars, measurements of the Doppler shift of the spectral lines providing estimates of the radial expansion/contraction velocity.

### *1.2. Stellar pulsations*

Many stars pulsate. Emblematic are the Cepheids which are found to pulsate in a well-defined region of the HR diagram, called the Instability Strip (IS). The mechanism that governs their pulsation is relatively well understood (Espinoza-Arancibia et al. 2023, Guzik et al. 2023 and references therein). As for most pulsating stars, it is a valve mechanism that controls the way by which heat can escape the star. In the case of Cepheids, where heat escapes mostly radiatively, it is called the kappa (and gamma) mechanisms: the opacity of the He II ionization zone increases under compression rather than decreases as expected from Kramer's law; the zone absorbs the energy released during compression for further ionisation rather than raising the temperature, thus increasing its opacity. The period-luminosity relation is so remarkably narrow that Cepheids can be used as clocks helping with measuring how far they are from us (Bono et al. 2024).

In the case of AGB stars, one speaks of Long Period Variables (LPV) including Mira, semi-regular (SRa and SRb) and irregular variables. A recent review by Skowron & Soszýnski (2026) gives a good overview and provides useful references. Miras are meant to pulsate regularly, allowing for the clear definition of a period and amplitude, the former exceeding 100 days and the latter exceeding 2.5 mag in the visible. Semi regulars are meant to have less clear a regime of pulsation, with amplitude smaller than 2.5 mag, the distinction between SRa and SRb referring to their proximity to Mira variables (SRa) or to irregular pulsators (SRb). However, in practice, this classification proves often to be inadequate: light curves may display regular pulsations of small amplitude, as is often the case for carbon-rich stars, or irregular pulsations of large amplitude: the oscillation amplitude is not a good proxy of the regularity of the pulsations. As a result, while qualitatively helpful, these distinctions are often ambiguous, or even misleading, and we shall refrain from making use of them.

Miras are located in cooler regions of the HR diagram than Cepheids are, in a less clearly delimited zone, roughly parallel to the IS. At variance with the IS, which is crossed nearly perpendicularly by the star path, the AGB instability region is run through from lower to larger luminosities, implying that Miras stay pulsating over a longer span of the HR diagram than Cepheids do. Attempts at relating the shape of their light curves in the visible to the state of their evolution along the AGB are hampered by the lack of a precise knowledge of the nature, location and morphology of the instability region in the HR diagram. The complexity of the mechanisms that govern pulsations has prevented this to be clarified. Indeed, while much progress has been made between early (Ostlie & Cox 1986) and present (Trabucchi & Pastorelli 2025) attempts at defining the edges of the instability region, we are still far from having a reliable picture.

The study of the regularity of the pulsation regime has revealed its complex nature. Irregularities in period have been the object of numerous studies and traditionally classified as meandering, continuous and sudden (Zijlstra & Bedding 2002). Such changes are often associated in the published literature with the possible recent occurrence of a thermal pulse. Indeed, a comprehensive modelling of the case of T UMi (Fadeyev 2022) has given evidence for the validity of such an interpretation in this specific case and many authors (Wood & Zarro 1981, Fadeyev 2022 and references therein) have produced detailed studies of the impact of thermal pulses on the appearance of a star. Irregularities concerning the luminosity at maximum light, the amplitude of oscillations and the shape of the light cycle have proven to be significant but poorly understood. They may take different forms; Hoai et al. (2026) have shown that fluctuations in maximum light and fluctuations in the shape of the ascending branch are uncorrelated, giving evidence for different types of irregularity. We know many examples of changes of amplitude, of shape, and even sometimes of period on a short time scale.

While the pulsation of Mira variables is believed to be caused by a valve mechanism, implying in particular partial ionization of the HI zone, its nature is far from being understood with the same reliability as in the case of Cepheids. The main difference is that heat escapes now dominantly convectively rather than radiatively, and, being cooler and larger than Cepheids, Miras have a weaker surface gravity and consequently experience important mass losses. Moreover, their convective envelopes are known to be structured in the form of convective cells, making an isotropic description inadequate, and to host strong shock waves. Their large radii imply long periods, ranging typically between 100 days and a few years. The regularity of their pulsation regime is much less than for Cepheids, with often significant differences between the shapes and scales of successive cycles. The relation between the light curve and the state of expansion/contraction is studied in a few cases and found to take the form of an anti-correlation, the star being brighter when warmer and smaller (see for example Trabucchi et al. 2023 and references therein). What is observed is the light emitted by the photosphere and atmosphere of the star and depends therefore on the nature of the atmosphere, both in terms of gas and dust. Moreover, when observing a single band-pass, the variation of the luminosity across a pulsation cycle is due both to the size and temperature of the star, as when observing the whole spectrum, and to the fact that the narrow window of the band-pass scans through different regions of the emitted spectrum. The latter contributes a positive correlation between the oscillation amplitude and the colour index, and explains the difference between the oscillation amplitudes in the visible and in the infrared (Reid & Goldston 2002).

Describing stellar pulsations as standing sound waves inside the star provides direct information about the star interior and has inspired numerous successful studies in the case of Cepheids, such as the description of the Hertzsprung progression in terms of a resonance between the fundamental and the second overtone (Marconi et al. 2024, Buchler et al. 2024 and references therein). In the case of AGB Mira variables, such a description is less productive; there has been a long debate in the last decade of the past century (Bedding et al. 1998) on their nature as fundamental or first overtone pulsators, the former being finally preferred and unanimously accepted at the end of the decade (Wood 2000). Much of the arguments of relevance rest on the observation of period-luminosity relations in the Magellanic clouds, for which the distance to us is known, allowing for a reliable measurement of the luminosity in the near infrared (Wood 2000). However, we know today from high resolution studies how complex is the dynamics governing the convective envelope: short time scale variations, convective cells, shock waves, mass ejections, etc... we cannot expect Fourier/wavelet analyses to explain what is going on as they do for organ pipes and violin strings (Hey et al. 2025, Chen et al. 2024, and references therein). In particular talking about multimode, mode switching, etc., for stars that do not pulsate clearly enough does not help with understanding what is going on. A very clear illustration of the complexity of the mechanisms at play is given by Ahmad et al. (2025) who compare state-of-the art current knowledge with the properties of three real pulsators, R Dor, RY Dra and AF Cyg. Fourier/wavelet analyses provide a parameterization of the light curve which needs to be compared with non-linear dynamical models (Buchler 1993) in order to gain understanding of the mechanisms at play. When the non-linear dynamics is dominated by only two modes, such a description is helpful and successfully describes effects associated with beats and resonances, such as the presence of humps or double-peaking, which are found in all pulsating systems.

To learn about stellar pulsations in the interior of a star from observations on its surface and surrounding atmosphere, one needs to build models. Most models assume radial pulsations; they start by having a hydrostatic description of the star interior obtained from what is known of stellar evolution. One then perturbs such an equilibrium state and observes its evolution: it may come back to equilibrium, in which case it is stable, or be unstable and evolve asymptotically to a regular pulsating regime, defining the pulsating state. One may use a linear or non-linear perturbation[1]. Examples are given by Bono et al. (1999) for Cepheids and by Fadeyev (2023) for AGB stars. In the case of convergence to a regular pulsating regime, one speaks of fundamental or overtone modes depending on the presence or absence of a node in the radial velocity distribution inside the star. Much progress has been achieved in producing reliable models of the pulsation of Long Period Variables. Recent examples are Ya'ari & Tuchman (1999), Fadeyev (2023), Trabucchi & Pastorelli (2025) and Ahmad et al. (2023, 2025). Yet, we are far from having a deep understanding of the relation between their light curves and the morpho-kinematics inside the star. Their study must accordingly be approached with an open mind, beyond the narrow frame of studies restricted to an analysis in terms of modes of pulsation; it must seek different ways of expressing the broad variety and irregularity of shapes displayed by their light curves. In particular, it is important to be conscious of the major oversimplification of the problem when viewed exclusively in terms of the period luminosity relation that they approximately obey (Wood & Sebo 1996).

In summary, there is a long way between observing a light curve and giving a reliable interpretation of what it corresponds to inside the star. However, whatever distortion it implies is believed to be continuous, changes in the light curve being closely related to changes in the state of the star interior. This is therefore encouraging efforts to seek relations between the shape of the light curve and the evolution of the star on the AGB and is motivating the present work.

## 2. Aim and method of the present work

In three preceding papers (Hoai et al. 2025, Paper I, Nhung et al. 2025, Paper II and Hoai et al. 2026, Paper III) we studied possible relations between the evolution of Mira variables on the AGB and the properties of their light curves. The justification was that we could expect reasonable continuity between both, with the exception of thermal pulses; however, the light curve depends not only on what happens inside the star, which is essentially a function of age and initial mass and metallicity, but also on what happens outside the star, which is essentially an unwelcome complication (dust, companions, etc...). The method consists in selecting stars for which both stellar parameters and light curve parameters have been evaluated and to try to reveal paths between them in both parameter spaces. It implies restricting the study to stars for which the light curve has been measured with sufficient quality and density of observations, which we do by selecting them from the AAVSO data-base in the visible.

We successfully identified light curve parameters such as the pulse width and, related to it, the shape of the ascending branch, as showing clear relations with the stellar parameters. However, a major problem that we met was the restriction of the studied sample to Mira variables: we could reveal paths in the region of instability where stars pulsate but we were unable to identify progenitors or successors of such stars when they are not pulsating with sufficient regularity to be part of the selected sample. The aim and motivation of the present paper is to overcome this deficiency.

As commented in the previous section, we need to free ourselves from categorisations such as Miras and semi-regulars, or as fundamental and overtones (which does not mean to ignore them) in order to approach the problem as pragmatically as possible, in a way adapted to the specificity of the data sample that we use, exclusively made of AAVSO observations in the visible. Having the possibility to follow the evolution of a light curve over a century, which is the exclusivity of light curve observations in the visible, shows the limitation of observations based on only a few cycles; however, their accuracy is limited, in practice preventing reliable results to be obtained when the oscillation amplitudes are smaller than some 0.2 mag. Moreover, the reliability of observations made with the naked eye by amateur astronomers may be a matter of concern; but whenever we had a chance to compare AAVSO observations with observations by professionals using sophisticated instrumentation, the agreement was always satisfactory, of course within the much larger uncertainties attached to the former (Lebzelter & Kiss 2001).

[1] We recall that a linear dynamical system is defined as being described by a differential equation of the form $dx/dt=Ax$ where $x$ is a vector function of $t$, $A$ a matrix, and $t$ the time. If $x_0$ is the initial state and $r_k$ and $\lambda_k$ the eigenvectors and eigenvalues of $A$, $Ar_k=\lambda_k r_k$, then the solution is $x=\Sigma(x_0.r_k)r_k exp(\lambda_k t)$.

The present work uses, in addition to the samples of Mira variables studied in Papers II and III, a sample of AGB stars which do not display such regular pulsations and for which AAVSO light curves are available with sufficient density and quality of observations. The stars, mostly classified as semi-regulars, are taken from Cadmus (2021, 2024), Percy & Tan (2013), Percy & Kojar (2013), Bergeat & Chevallier (2005), Lebzelter & Hron (1999), Darriulat et al. (2024) and Kiss et al. (1999). The whole sample of 245 stars is listed in Table A1 of the Appendix, including detailed information on their stellar and light curve parameters as described below.

Spectral types are taken from the SIMBAD data base and the colour index *C*, generally defined as *K*-[22], is taken from Cutri et al. (2003, 2021) and Schlafly et al. (2019).

An estimate of the mean amplitude of the oscillations, *A*, measured in magnitudes, is given for stars displaying some evidence for pulsation. For the others, *A* is a measure of the mean variation of the magnitude over times of short duration, typically one year. This definition differs from that used by several authors in the published literature. For example, we compared our values of *A* to those assigned by Glass & Van Leeuwen (2007) to a sample of 29 stars, which we both studied. Major disagreements were observed, with our values being occasionally much smaller because Glass & Van Leeuwen (2007) define *A* as the difference between the maximal and minimal values over the whole light curve, which is essentially meaningless in terms of understanding pulsations. Yet, the values quoted for *A* in Table A1 are not very meaningful either when the mean value of *A* varies strongly over the sample.

The column labelled ‘Ref’ in Table A1 lists some most instructive and/or recent articles of relevance to the star. In particular, they are the source of the values quoted for the effective temperature, *T* (K), the luminosity *L* (measured in units of 1000 solar luminosity), the radius *R* of the stellar disc (measured in solar radii), the mass-loss rate $\dot{M}$ (measured in units of $10^{-7}$ solar masses per year), the presence or absence of technetium in the spectrum and the $^{12}C/^{13}C$ isotopic ratio. When different values of a same quantity are quoted by different authors, the table lists an estimate of their mean. Very large uncertainties may affect these parameters, but their values are meant to provide sufficient information for the arguments developed in the text of the article to apply; however, they should not be quoted, the only proper sources being those listed in the column ‘Ref’.

For light curves displaying clear pulsations, we have evaluated the mean value of the period (measured in days), but we refrained from doing so otherwise, the values obtained in such cases failing to provide useful information, or, worse, being misleading. In several cases, such as Long Secondary Periods (LSP) or double-maxima, defining a period is not obvious, in which cases we discuss the issue in the text. Whenever permitted by the quality of the light curve, we have evaluated the shape parameters *W* and *q*, which were defined in Papers II and III, respectively. We recall that *W* is a measure of the duration of the light pulse relative to the period and *q* a measure of the deviation of the shape of the ascending branch from that of a sine wave (a sine wave in luminosity has $W$~0.6 while a sine wave in magnitude has *W* between 0.3 and 0.5 and $q$=0).

The first step of the analysis was to carefully scrutinize, one by one and over the whole past century, each of the light curves of the Table A1 sample. Some were found to display regular pulsations, some to pulsate only episodically and some to essentially not pulsate at all. It then proceeded, by trial and error, to group together stars for which both the stellar and light curve parameters were closely similar in order to form families playing the role of meaningful milestones along the paths of evolution of both stellar and light curve parameters. Several iterations were necessary before reaching a state in which a plausible picture had emerged. We do not report about this preliminary work in the present article but use the plausible final picture as a guide to present the results of our work, discussing carefully the strong and weak points of each separate step implied by the suggested evolution scheme. This is done in the next section.

## 3. A suggested scheme of evolution

The detailed scrutiny of the light curves of some 250 stars, including those studied in Papers II and III and the sample of semi-regular and irregular stars introduced in the present article, in relation with the properties of the stellar parameters, suggests a scheme of their evolution along the AGB, which we introduce in the present section. Figure 2 summarizes its main features, illustrated by families of stars representative of the main milestones of the suggested evolution. The mean and rms values of the stellar and light curve parameters of each family are listed in Table 1, and Table A1 of the appendix lists each family separately.

Some stars enter the AGB already pulsating (family A), others not (family B).

Family A1 (Table 2 of warm Mno Mira variables in Paper III) collects a rather compact group of early spectral type stars which pulsate regularly with large amplitudes and relatively short periods, implying that their progenitors must have spectral types earlier than M4 and pulsate with periods smaller than 200 days. We know very few examples of such stars having reached the AGB, we collect a few examples in family A; this implies that some progenitors of family A1 had not yet entered the AGB but were already pulsating. If not evolving to family A1, the family A stars may also, when having large enough initial masses, become Hot Bottom Burners, forming the HBB family. The stars of the A1 family will cool and grow, entering the family Ano of the Tc-poor stars that were studied and called ‘Mno’ in Paper III. They may then either keep ageing without experiencing strong TDU events and become post-AGB stars or experience strong TDU events to become Tc-rich stars that were studied and called ‘My1’ (or ‘S1’) in Paper III, forming the Ayes family. These will either leave the AGB as oxygen-rich or become carbon-rich.

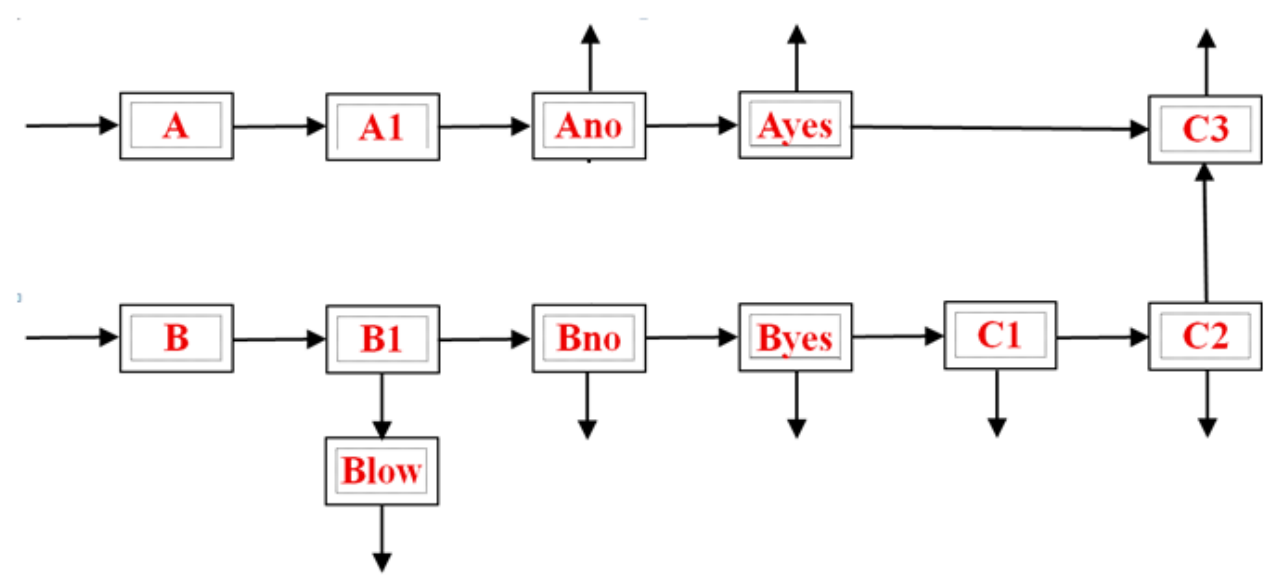

**Figure 2.** Schematic of the suggested evolution of the progenies of the A and B families (see text).

The stars that enter the AGB not yet pulsating[2] do it with amplitudes not exceeding 0.4 mag and colour indices not exceeding 0.4 mag; they are the earlier members of family B. Some may keep aging as non-pulsating low colour index and low amplitude stars and ultimately leave the AGB, but most will acquire colour and/or amplitude before reaching spectral type M6 with amplitudes not exceeding 0.6 mag and evolve to family B1, made of low amplitude spectral type M5 stars. Some of these will keep ageing beyond spectral type M6 and are likely to have low initial masses and leave the AGB before experiencing strong TDU events; they form family Blow. The others will acquire larger amplitudes and eventually enter the TPAGB, forming successively the Bno and Byes families (groups 'My2', 'S2' and 'S3' of Paper III). They may leave the AGB before or after experiencing strong TDU events, and before or after becoming carbon-rich.

Carbon-rich stars have on average much lower amplitudes than oxygen-rich stars and, as a result, often seem not to pulsate. Many have amplitudes smaller than 1.2 mag, mostly smaller than 0.6 mag, and colour indices smaller than 2: they form family C1. The others form two families of different effective temperatures. The warmer stars, with $T$>~2400 K, are too warm to be successors of the cooler Ayes family and form family C2 while the cooler stars, with $T$<~2400 K form family C3.

**Table 1.** Mean/rms values of the stellar and light curve parameters associated with each family described in the text. Type is the spectral-type taken from the SIMBAD data-base; $C$, the colour index measured in magnitudes; $A$, the amplitude measured in magnitudes; $T$, the effective temperature measured in Kelvins; $L$ the luminosity measured in units of 1000 solar luminosities; $R$ the stellar radius measured in units of solar radii; $\dot{M}$, the mass-loss rate measured in units of $10^{-7}$ solar masses per year; $^{12}C/^{13}C$, the carbon isotopic ratio; $P$, the period measured in days; $W$ and $q$, the light curve parameters defined in Papers II and III, respectively. Tc=0 for Tc-poor and =1 for Tc-rich. When the values are taken from the published literature, they are usually affected by large uncertainties, and different authors often quote different values. Accordingly, the values listed in the table are estimates well adapted to the requirements implied by their use in the present work but not to be quoted: the only values to be quoted are from the sources listed in the column 'Ref' of Table A1, and references therein.

| Family | | Type | $C$ | $A$ | $T$ | $L$ | $R$ | $\dot{M}$ | Tc | $^{12}C/^{13}C$ | $P$ | $W$ | $q$ |
|---|---|---|---|---|---|---|---|---|---|---|---|---|---|
| A | 1 | 1.8/1.3 | 1.8/0.5 | 2.1/0.5 | - | - | - | - | - | - | 152/32 | 46/3 | -6/22 |
| A1 | 2 | 2.9/0.7 | 1.6/0.2 | 4.5/0.4 | 3450/50 | - | 180/50 | - | 0/0 | - | 186/32 | 35/3 | -47/20 |
| HBB | 3 | 4.8/1.2 | 2.2/0.6 | 3.1/1.8 | 3160/110 | - | - | 1.9/0.6 | 0/0 | 8/3 | 467/60 | 45/8 | -157/14 |
| Ano | 4 | 6.1/1.5 | 2.3/0.6 | 5.0/0.5 | 2730/230 | 5.2/1.0 | 340/12 | 4/4 | 0/0 | 13/4 | 355/83 | 25/5 | 76/69 |
| Ayes | 5 | 6.6/1.0 | 2.4/0.8 | 6.0/1.0 | 2340/410 | 7.0/1.0 | 560/190 | 11/10 | 1/0 | 29/7 | 417/70 | 23/3 | 64/43 |
| B | 6 | 3.0/1.4 | 0.5/0.3 | 0.3/0.1 | 3470/110 | 2.5/0.7 | 120/25 | 0.3/0.3 | 0/0 | 11/3 | - | - | - |
| B1 | 7 | 5/0 | 1.1/0.5 | 0.4/0.1 | - | 3.4/1.2 | - | 2.6/2.5 | 0/0 | - | - | - | - |
| Blow | 8 | 6.6/0.6 | 1.3/0.5 | 0.5/0.1 | 3070/260 | 3.5/1.2 | 230/60 | 2.3/1.6 | 0.4/0.5 | 16/7 | 186/98 | - | - |
| Bno | 9 | 4.9/1.2 | 1.6/0.5 | 1.2/0.5 | 3280/180 | 3.8/2.1 | 195/51 | 0.7/0.7 | 0.2/0.4 | 24/14 | 195/86 | 50/4 | -10/40 |
| Byes | 10 | 6.4/1.0 | 1.4/0.6 | 4.2/1.0 | 2730/340 | 6.6/1.3 | - | 5.2/7.4 | 0.9/0.3 | 27/8 | 326/64 | 37/6 | -71/49 |
| C1 | 11 | - | 1.4/0.3 | 0.4/0.2 | 2770/220 | 6.1/2.8 | - | 3.8/2.7 | 0.7/0.5 | 44/26 | 307/63 | - | - |
| C2 | 12 | - | 1.7/0.5 | 2.5/0.6 | 2860/190 | 5.0/1.4 | - | 9/9 | 1/0 | 17/11 | 397/61 | 47/8 | -57/51 |
| C3 | 13 | - | 2.6/0.6 | 2.9/0.6 | 2120/170 | 9/4 | - | 30/15 | 1/0 | 50/64 | 441/72 | 39/5 | 40/17 |

### *3.1 Family A*

Family A (Figure 3) collects examples of oxygen-rich stars having spectral types earlier than M4 and pulsating with periods not exceeding 200 days, therefore being possible progenitors of the warm Tc-poor Mira stars of family A1. Their mean spectral type is M1.8, some may be of spectral type K, their mean period is 152 days, their mean colour index is 1.8 and their mean amplitude is 2.1 mag. They stand out in the sample of early spectral types as having larger amplitudes. None of their basic stellar parameters is explicitly quoted in the published literature and only one star has been searched for technetium; it has been found Tc-poor. The light curves are close to sine waves with $\langle W \rangle$=0.46 and $\langle q \rangle$=-0.006. While their periods and colour indices are in continuity with those of their successors, families A1 and Ano, $q$ is not: the deviation from a sine wave proceeds by continuous decrease of $W$ but with $q$ reaching a minimum for the A1 family (Figure 3). In general, they pulsate regularly, RY Leo and X Lib being the least regular.

### *3.2 Family A1*

Family A1 collects the nine warm 'Mno' stars listed in Table 2 of Paper III together with the semi-regular RS Aqr (Figure 3). They are most likely successors of family A and progenitors of family Ano; they form a compact group with a mean spectral type M2.9, a mean colour index of 1.6 mag, a mean amplitude of 4.5 mag and a mean period of 186 days. Their large amplitudes illustrate how fast these increase in the early part of the instability region. The shapes of the light curves move farther away from sine waves with <*W*>=0.35 and <*q*>=-0.047, significantly smaller than those of their family A progenitors. The evolution from the A1 to Ano families is illustrated in Figure 4, showing how *q* starts becoming negative in the A to A1 transition and reverses its trend in the A1 to Ano transition. Whenever searched for technetium the A1 stars are found to be Tc-poor, except for two cases, one where the presence of technetium is described as "doubtful" and the other (R Vir) for which conflicting results have been published. Their successors are the other Tc-poor Mira variables studied in Paper III, collected in the Ano family and having larger periods, later spectral types, smaller values of *W* and larger values of *q*.

[2] We recall that in the present article, when we say that a star does not pulsate, it means that pulsations cannot be reliably evidenced from the AAVSO light curve. However, more sensitive observations may very well give evidence for clear pulsations.

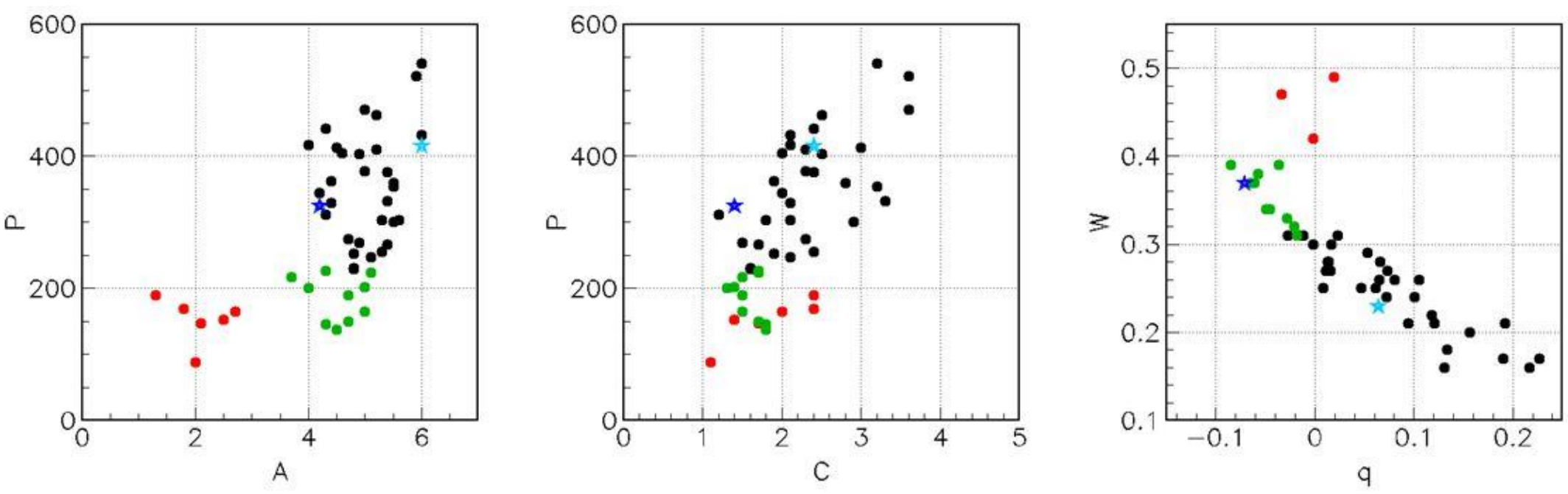

**Figure 3.** The A (red)→A1 (green)→Ano (black) evolution in the *P* vs *A* (left), *P* vs *C* (centre) and *W* vs *q* (right) planes. Stars show the mean locations of the Ayes (cyan) and Byes (blue) families.

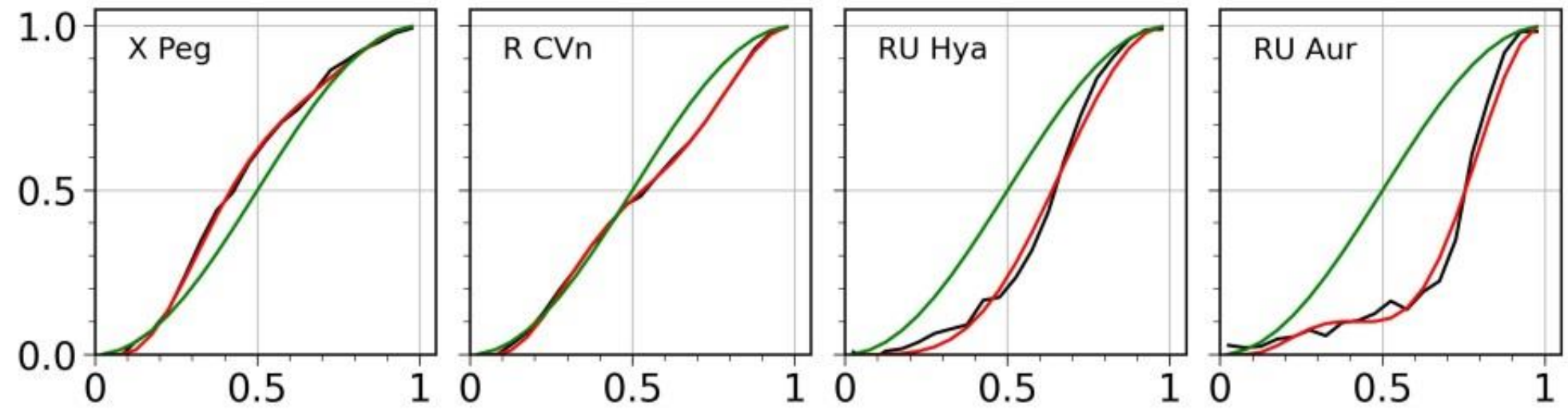

**Figure 4.** Deviation from a sine-wave shape of the ascending branches of family A1 (X Peg, *W*=0.39, *q*=-0.085) and family Ano (R CVn, *W*=0.30, *q*=-0.002, RU Hya, *W*=0.24, *q*=0.101 and RU Aur, *W*=0.21, *q*=0.192). In each panel, the green curve shows a sine wave.

### *3.3 Family HBB*

Hot Bottom Burners have initial masses larger than ~4 solar masses and their spectra include no technetium but heavy lithium lines. Family HBB collects seven possible candidates, of which six were discussed in Papers II and/or III; the other, WZ Cas, has been studied by Lebzelter et al. (2005), suggesting that in spite of its being carbon-rich its main properties are evidence for HBB. Their light curves display characteristic double-maxima. They most likely belong to the progeny of the A family.

### *3.4 Family Ano*

The Ano family collects the Mira stars studied in Paper III, likely progeny of family A1, having spectral types later than M3 and having not yet experienced a strong TDU event. Figure 5 displays distributions of the stellar and light curve parameters. In fact, out of these 31 stars only 19 have been shown to be Tc-poor; they are shown in red in Figure 5. Three stars stand out as having a larger period than expected from the value of their spectral type index, V Cam (M7, 521 d), X Cep (M4, 540 d) and R Tel (M5, 462 d); they have not been searched for technetium and are shown in green in Figure 5. They have large positive *q* values, suggesting that if they had experienced strong TDU events, they would be in the Ayes rather than Byes family: it would modify the sharing of stars between the Ano and Ayes families, which succeed each other, but would not modify the scheme adopted to describe the evolution. The mean values of the parameters of the Ano family are: M6.1, <*C*>=2.3 mag, <*A*>=5.0 mag, <*P*>=355 days, <*W*>=0.25, <*q*>=0.076. When measured, the $^{12}C/^{13}C$ isotopic ratio is always well below 20 (13 on average), consistent with the absence of strong TDU event. The mean effective temperature is 2730 K and the mean mass-loss rate 4 $10^{-7}$ solar masses per year. The progenitors of the Ano family are the stars of the A1 family (Figure 3). Strictly speaking, one cannot exclude that some stars from the B1 or Bno families might evolve to Ano, but this seems very unlikely from their location in the parameter space. Some Ano stars may never experience strong TDU events and simply keep cooling down and ultimately leave the AGB. The others, after having experienced strong TDU events will mostly evolve to the Ayes rather than Byes family, as implied by their location in the parameter space (Figure 3).

### *3.5 Family Ayes*

The Ayes family collects oxygen-rich Mira stars studied in Paper III as progeny of the Ano family and shown (11 stars), or suspected (5 stars), to have experienced strong TDU events. We added to these the Mira variable RV Peg. Of the latter five, three are of S spectral type and can therefore be safely assumed to be Tc-rich. The remaining two, RW And and S Pic, would fit well in family Ano if they have not experienced strong TDU events, therefore with no impact on the adopted scheme of evolution. The mean spectral types are M5.8 and S5.3 and the mean values of the other parameters are <*C*>=2.4 mag, <*A*>=6.0 mag, <$P$>=417 days, <$W$>=0.23, <$q$>=0.064. When measured, the $^{12}C/^{13}C$ isotopic ratio is well above 20, 29 on average, with the exception of R Ser, with a ratio of only 14; Hinkle (2016) evoked the possibility of it being a massive star for which HBB regulates the carbon isotopic ratio toward lower values but did not retain it as reliable. Some Ayes stars are expected to become carbon-rich, but others may remain oxygen-rich and ultimately leave the AGB.

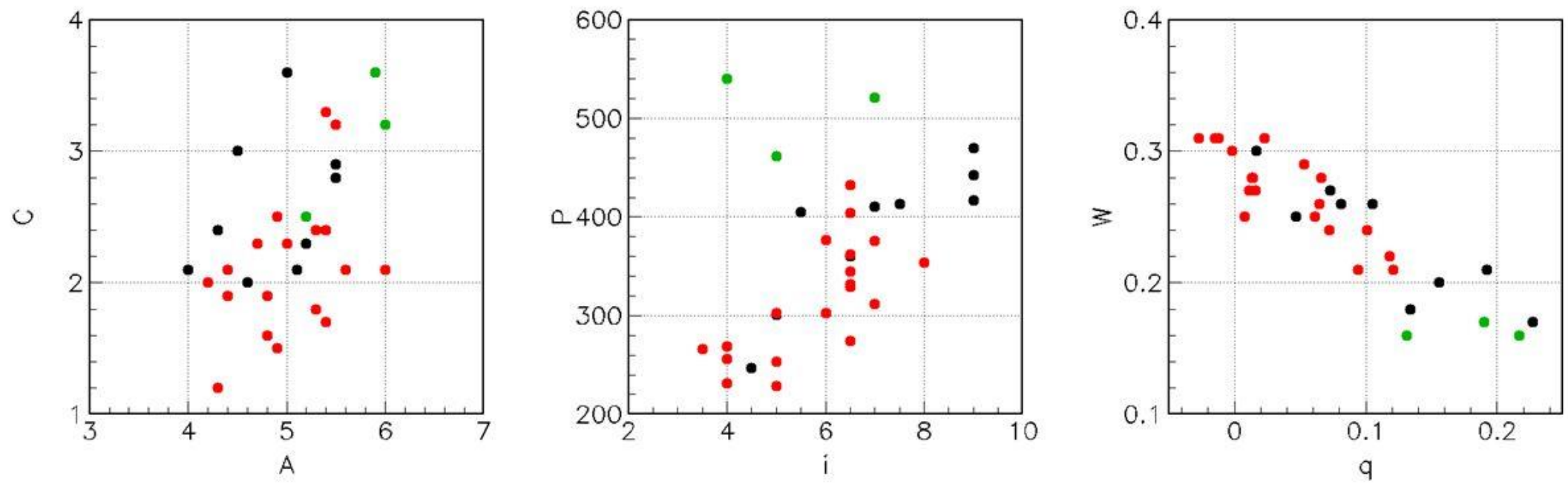

**Figure 5.** Distributions of the stellar and light curve parameters of family Ano in the $C$ vs $A$ (left), $P$ vs $i$ (centre) and $W$ vs $q$ (right) planes. The stars that have been found to be Tc-poor are shown in red. The stars shown in green (V Cam, X Cep, R Tel) are discussed in the text.

Figure 6 displays distributions of the stellar and light curve parameters. T Sgr was discussed in Paper III, the low value of its amplitude being blamed on the presence of an unresolved F companion. U Cas, with low values of both colour index and period, would better fit in the Ano family if it were not for its S spectral type, which, however, is reliably evaluated. RV Peg, which stands a bit out in the $C$ vs $A$ plane, displays a clear Long Secondary Period and is discussed in Section 4.

Stars R Aur and chi Cyg stand a bit out in the $W$ vs $q$ plane as having low $W$ and/or $q$ values, but follow the average trend in the other planes; the light curve of R Aur was discussed in Paper III as having a very low value of parameter $p$. In the case of chi Cyg, both $W$ and $q$ values are reliably measured, the very large amplitude being the main peculiarity; the star has been abundantly studied (Hoai et al. 2024 and references therein), found to be in a state of high instability with anisotropic mass-loss varying on a short time scale and with a mass of ~2.1 solar masses (Lacour et al. 2009), implying a large initial mass nearing 3 solar masses.

S Cas stands out as having a large period (613 days) and a low temperature (1800 K). Danilovich et al. (2015) note its large mass-loss rate and terminal expansion velocity, its standing on the higher C/O end of the S star scale and its bearing significant similarities with carbon stars. It is claimed to have a large mass of 4.1 solar masses by Khalatyan et al. (2024).

In principle, the progenitors of Ayes stars may be from the Ano or Bno family, as discussed in Subsection 3.10 after having introduced the Bno and Byes families. But in both cases, they can only be from the earlier members of the Ano or Bno families, with the spectral type index $i$ not exceeding 5, implying that the Ano stars having $i$>5 are likely to keep cooling and expanding as Tc-poor stars and to leave the AGB without experiencing strong TDU events. Some of the Ayes stars will become carbon-rich, others will leave the AGB as oxygen-rich.

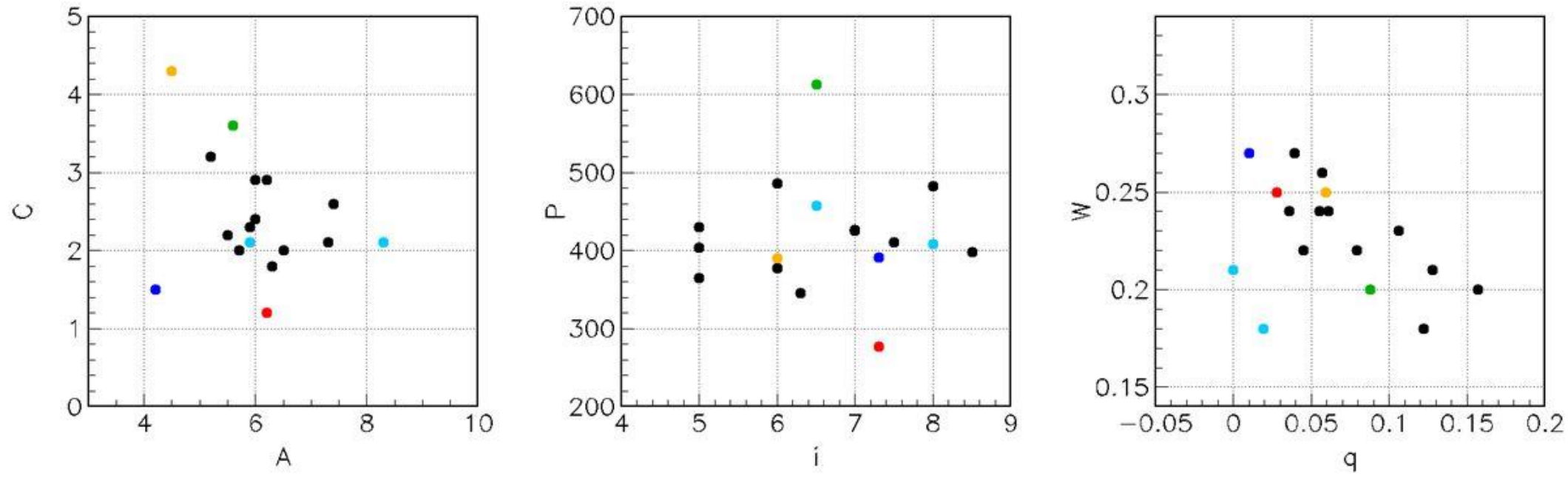

**Figure 6.** Distributions of the stellar and light curve parameters of family Ayes in the $C$ vs $A$ (left), $P$ vs $i$ (centre) and $W$ vs $q$ (right) planes. In the case of stars of S$j$ spectral type, the spectral type index plotted in the central panel is $i$=5+$j$/2, an approximation matching the M$i$ scale of spectral types. R Aur and chi Cyg are shown in cyan, T Sgr in blue, U Cas in red, S Cas in green and RV Peg in orange (see text).

### *3.6 Family B*

The B family collects stars entering the AGB without significantly pulsating. Its progeny is illustrated in Figure 7. In terms of modes of oscillation, in contrast with the A progeny, which is dominated by Mira fundamental pulsators, the B progeny is dominated by overtone pulsators, which is, however, meaningless if the star does not pulsate. Figure 7 suggests several possible evolutions within the B progeny, schematically indicated by arrows: the star may leave the AGB without having significantly started to pulsate, or with having started to pulsate and to lose mass but before experiencing strong TDU events, or after having experienced strong TDU events but before becoming carbon-rich, or after having become carbon-rich. In order to organize the discussion, we divide the B progeny in families illustrating the main milestones of the suggested possible evolutions. These are purely indicative, the boundaries between them are only crudely defined, they should not be thought of as strictly different categories, but as milestones on continuous evolution paths.

The stars of the B family are oxygen-rich with early spectral types, M3.0 on average, small amplitudes, 0.3 mag on average and small colour indices, 0.5 mag on average. They are warm, with a mean effective temperature of 3470 K and, whenever they have been searched for technetium, they have been found Tc-poor. They have had no time to build a significant circumstellar envelope, have negligible mass-loss rates and are barely variable. When measured, the $^{12}C/^{13}C$ isotopic ratio is well below 20, 11 on average, typical of stars that have not yet experienced strong TDU events. While some may in principle cool down and ultimately leave the AGB without significant increase of their amplitude and colour index, most will start such evolution before reaching spectral type M6; they will enter directly the Bno family if they leave the B family before reaching spectral type M5; otherwise, they will keep cooling down to spectral type M5 and enter the B1 family.

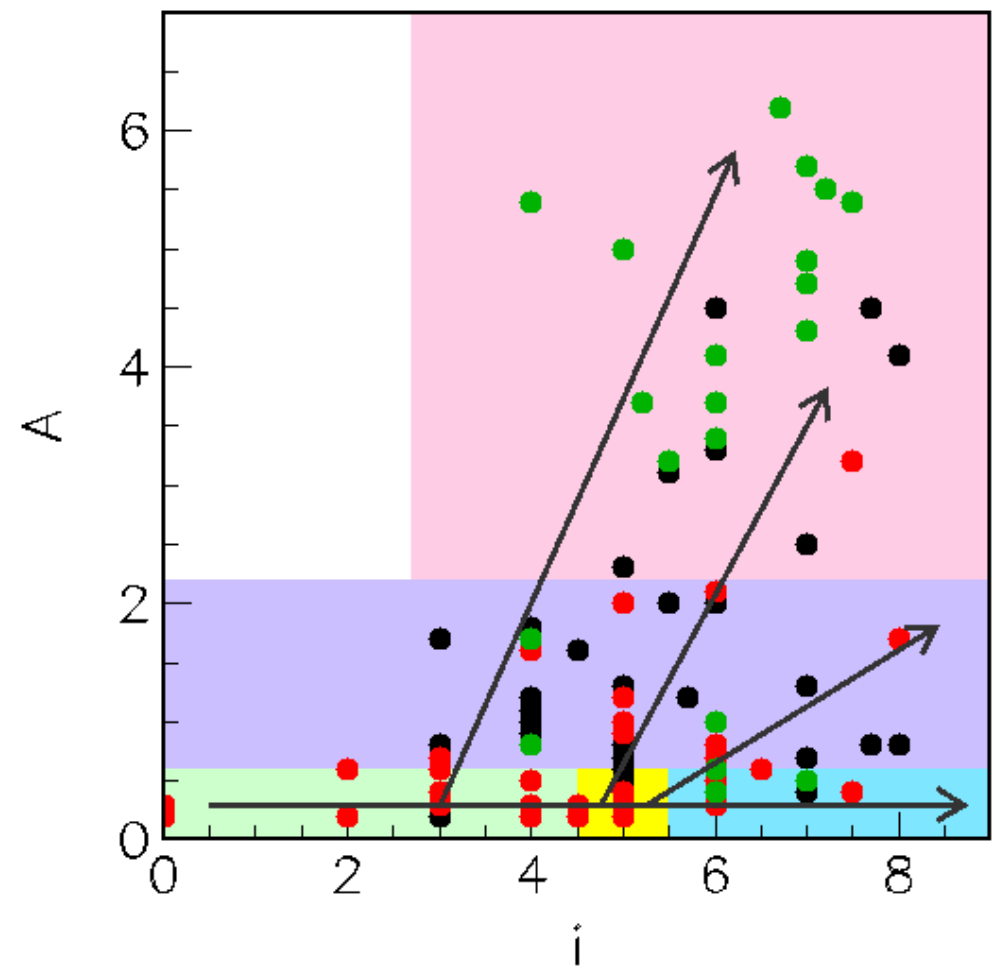

**Figure 7.** Evolution in the *A* vs *i* plane of family B and its progeny. Tc-rich stars are shown in green and Tc-poor stars in red. Different families are coloured in green (B), yellow (B1), blue (Blow), purple (Bno) and red (Byes). Arrows illustrate schematically possible paths of evolution.

### *3.7 Family B1*

The B1 family collects a few stars of spectral type M5 that have family B stars as progenitors; it acts as a switch leading either to family Bno, in which case the stars start to pulsate with larger amplitudes, or, for stars having small enough initial masses, to family Blow, in which case they will cool down and ultimately leave the AGB with low amplitudes. The B1 stars are still essentially non-pulsating, with an amplitude never exceeding 0.6 mag, 0.4 on average. They have a mean colour index of 1.1 and are of course Tc-poor with small mass-loss rates.

### *3.8 Family Blow*

These are stars that succeed family B1 without acquiring significant amplitude and keep cooling down. As we know that AGB stars having a low initial mass, below ~1.5 solar masses, are expected to leave the AGB prematurely before experiencing strong TDU events, it is tempting to identify family Blow with candidates for such an evolution. They would become post-AGB stars that are not s-process enriched (Kamath 2020, Kamath & Van Winckel 2022). The case of R Dor speaks in favour of such an interpretation, with an initial mass reliably estimated to be between 1 and 1.5 solar masses by Ohnaka et al. (2019); a similar result is found for RX Boo (Kamezaki et al. 2012) and theta Aps (Ohnaka et al. 2014); all three stars are Tc-poor with temperatures in the 2700 K range. But other stars in the family are significantly warmer and three of them are Tc-rich: OP Her and ST Her, with an MS spectral type, and SW Vir. On average, they have cooled down to an effective temperature of 3070 K without acquiring much colour, $<C>$=1.3 mag. The spectral types are later than M5, M6.6 on average, and the amplitudes never exceed 0.8 mag, 0.5 on average. When measured, their $^{12}C/^{13}C$ isotopic ratios and mass-loss rates are small, 16 and 2.3 $10^{-7}$ solar masses per year on average, respectively, and when searched for technetium, they have been generally found to be Tc-poor. The periods range between ~75 and ~350 days, with an average value of ~195 days.

### *3.9 Family Bno*

In contrast with the Blow family, the Bno stars are successors of B and B1 stars that acquire significant amplitude when ascending the AGB, 1.2 mag on average, large enough to allow for a reliable estimate of the pulsation period, measured to be 195 days on average. They are still warm, 3280 K on average and early, with a mean spectral type of M4.9. When searched for technetium, they are usually found to be Tc-poor, with the exception of RS Cnc, U Del and W Cyg. Their colour index has only slightly increased, 1.6 mag on average, and their mass-loss rate is still low, at the level of $10^{-7}$ solar masses per year. The shape of their light curves is close to sine wave, with $<W>$=0.50 and $<q>$=-0.010. Figure 8 displays their distribution in a few planes of the parameter space.

The period is seen to increase as a function of the spectral type index, very approximately as 50($i$-1) days for the majority but as twice this value for RU Per, W Tau, RY UMa and RW Boo and as half this value for AF Cyg, TX Dra, g Her, X Her and S Lep. While simply described as different oscillation modes in terms of the period-luminosity relations (Bedding & Zijlstra 1998), an interpretation in terms of the mechanisms at play is less straightforward. Of the five low period stars, one, AF Cyg, is a clear case of double period discussed in Subsection 4.2 and the four others, g Her and S Lep very clearly, TX Dra and X Her somewhat less, together with two M4 stars of the same family, V465 Cas and U Del, are cases of Long Secondary Periods discussed in Section 4.

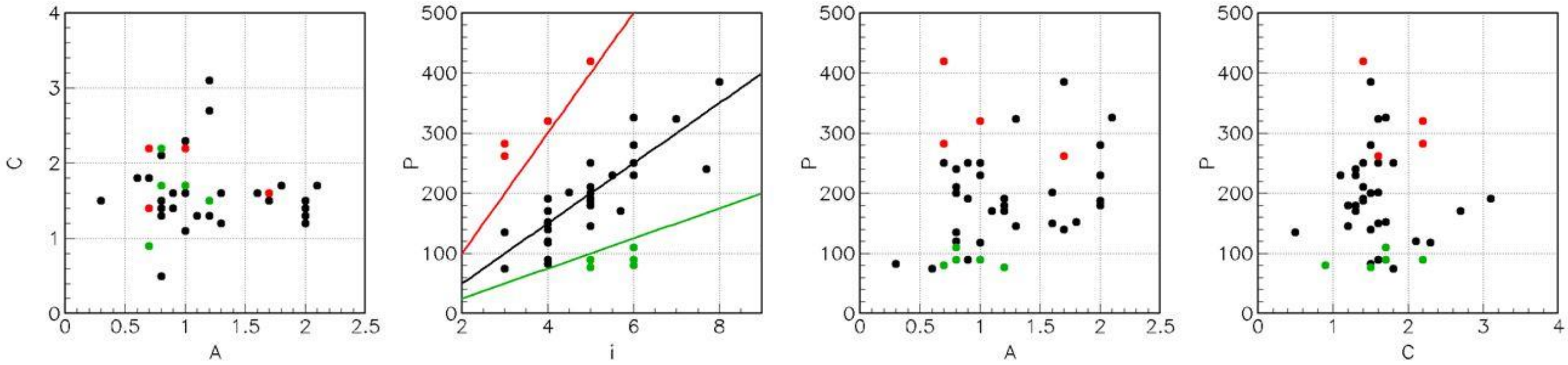

**Figure 8.** Distribution of the stars of family Bno in various planes of the parameter space: from left to right, $C$ vs $A$, $P$ vs $i$, $P$ vs $A$ and $P$ vs $C$. The stars with large periods are shown in red and those with small periods in green (see text). In the $P$ vs $i$ panel the lines $P$=25($i$-1) (green), 50($i$-1) (black) and 100($i$-1) (red) are drawn as references (see text).

Among the large period cases, RU Per stands out as failing to display reliable data; the period quoted in Table 1, 320 days, corresponds to a few oscillations in the 1930s and 1970s; but the light curve is otherwise very irregular, occasionally displaying different periods, such that Percy & Tan (2013) quote values of 93 and 602 days and the General Catalogue of Variable Stars quotes a value of 170 days. RW Boo pulsates with a period of ~420 days between 1984 and 1998 with an amplitude of 0.4 mag and between 2013 and 2018 with an amplitude reaching 1 mag but otherwise essentially does not pulsate; Alonzo Hernandez et al. (2024) give evidence for two wind velocities, 8 and 16 km/s and quote a mass loss rate of 3.5 $10^{-7}$ solar masses per year. W Tau and RY UMa have not been the objects of specific studies in the published literature but the light curves display good density and quality of observations. That of W Tau, measured by Cadmus (2024) displays two clear peaks in its Fourier spectrum, one with a period of 244 days, and the other, about half in amplitude, with a period of 128 days. RY UMa usually pulsates with a period of 282 days and an amplitude of ~0.6 mag but displays three 5-year-long episodes, centred in1973, 1986 and 2000, during which it stops pulsating.

In all cases, the stars which stand out in the $P$ vs $i$ plane have values of $C$ and $A$ consistent with the mean values of the family.

Some of the Bno stars may end their AGB lives with such moderate amplitudes while others will possibly experience strong TDU events, acquire larger amplitudes and enter the Byes family.

### *3.10 Family Byes*

Family Byes collects the successors of family Bno that have acquired larger amplitudes, 4.2 mag on average, and are likely to experience strong TDU events: out of 16 stars that have been searched for technetium, 14 have been found to be Tc-rich; the other two are V Boo, which was first found Tc-rich and later claimed to be Tc-poor, and W Hya, which has been abundantly studied in the literature (Ohnaka 2025 and references therein) and found difficult to understand; note that the relatively small amplitude oscillations of its light curve (~3.2 mag) contrast with the relatively large amplitude oscillations of the radial velocity (~15 km/s, Lebzelter 2025). Star GY Aql stands out as having a very large mass-loss rate, an order of magnitude larger than the mean rate of the family (Wallström et al. 2024); it is a privileged target of the ATOMIUM collaboration, who however fail to produce a reliable explanation of the large mass-loss rate (Pimpanuwat et al. 2026 and references therein); the light curve is unfortunately poorly measured.

In principle, Ayes and Byes families might have Ano and/or Bno progenitors, but in practice a transition from Ano to Byes is unlikely. Figures 3 and 9 illustrate this point by comparing the Ano and Bno evolutions in the $P$ vs $C$, $W$ vs $q$ and $P$ vs $i$ planes, in relation with the regions covered by the Ayes and Byes families. Both the Ayes and Byes families have spectral types in the neighbourhood of M6.5 and $^{12}$C/$^{13}$C isotopic ratio having reached ~30; but the Byes family shows signs of being less evolved than the Ayes family, with a higher mean temperature (2730 compared with 2340 K), a lower mean period (326 compared with 417 days), a smaller mass-loss rate (5.2 compared with 11 $10^{-7}$ solar masses per year) and significantly broader pulses ($\langle W \rangle$=0.37 instead of 0.23 and $\langle q \rangle$=-0.071 instead of +0.064). This makes transitions between the A and B progenies rather unlikely.

Figure 10 displays distributions of the Byes family in a few planes of the parameter space. The semi-regular Byes stars, which have a light curve sufficiently well measured for the shape parameters $W$ and $q$ to be evaluated, V Boo, X Oct and U Per, are shown in red in Figure 10, intermediate between the mean Bno family and the Byes Mira variables of group S3 of Paper III, R Cam, T Cam, T Gem and S UMa, which reach very negative $q$ values. This gives clear evidence for an evolution from the Bno family to the stars of this S3 group of Paper III as schematically shown with a blue arrow in Figure 10.

The remaining Byes stars include the six My2 stars of Paper III, T Cep, S CMi, S Her, R Hya, Z Peg, and U UMi and the four S2 stars of Paper III, V Cnc, Z Del, R Gem, and R Lyn. In addition, they include two Mira variables, FF Cyg and T Cas, which were shown in Paper III to sit between the groups My2 and S3 and three semi-regular variables, GY Aql, RZ Cyg and W Hya. They form a compact isolated group in the $W$ vs $q$ plane, apparently shedding doubt on their being the progeny of the Bno family. The answer to this puzzle is simple: the stars making the transition between them and the Bno family are members of the Bno family rather than members of the Byes family. The low amplitude of the Bno stars allows for only a few of them to have their light curve parameters reliably measured, but among these there are three clear candidates for the missing transition: U Boo, X Mon and Z UMa; they are shown in purple in Figure 10.

Like in the case of the Ayes family, some Byes stars will end their lives on the AGB as oxygen-rich while some others will become carbon-rich.

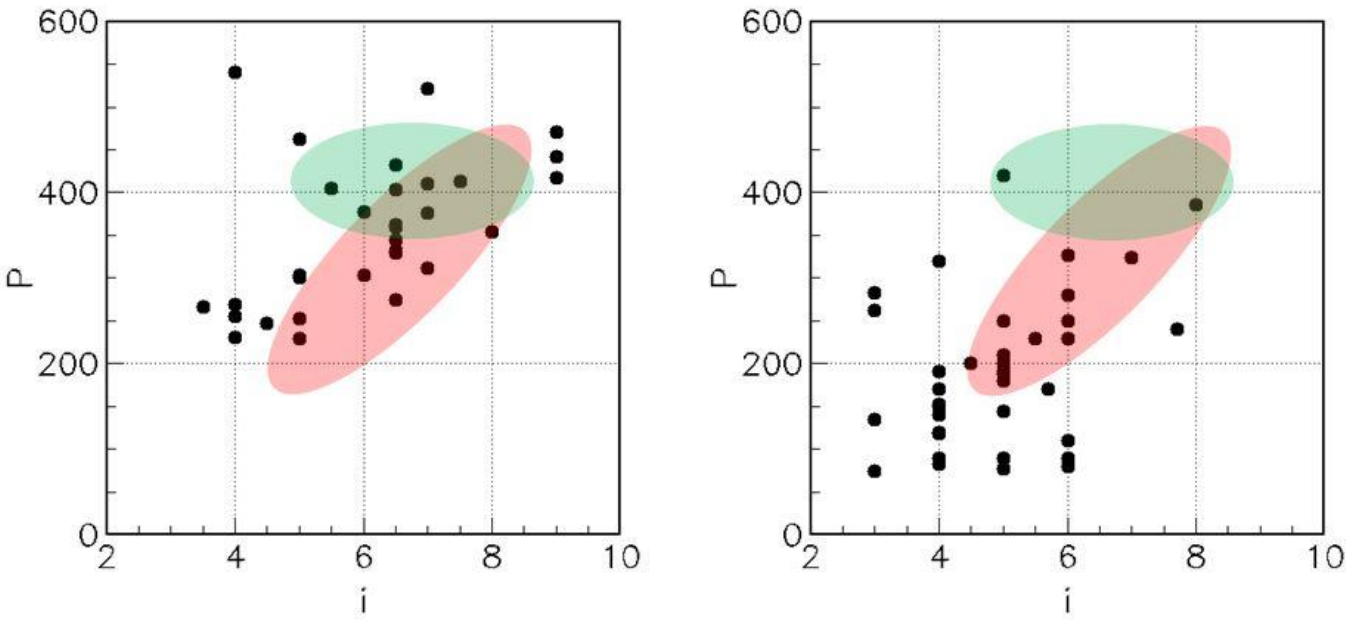

**Figure 9.** $P$ vs $i$ distributions of the stars of families Ano (left) and Bno (right). The ellipses show the approximate locations of the Ayes (green) and Byes (red) families.

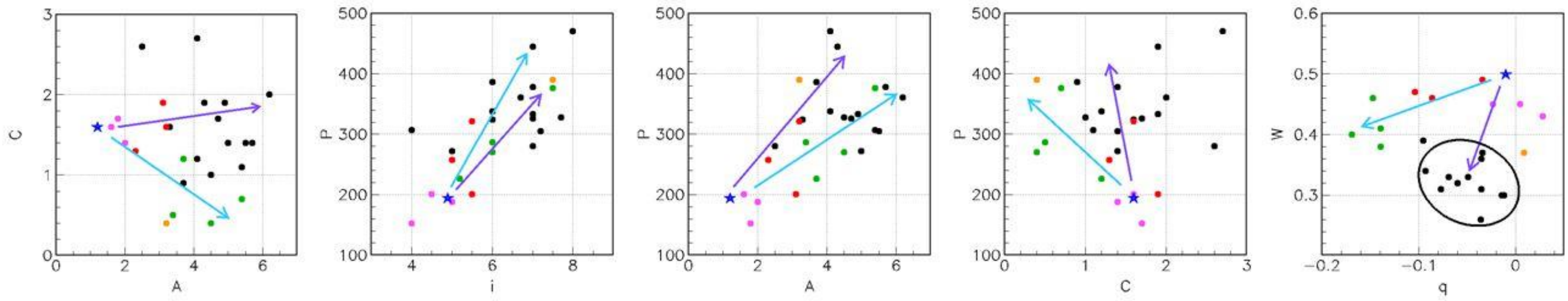

**Figure 10.** Distribution of the stars of family Byes in various planes of the parameter space. The blue star shows the mean location of the Bno family. The Mira variables of the S3 group of Paper III are shown in green; the three semi-regular variables V Boo, X Oct and U Per, making the transition to the Bno family, are shown in red. W Hya is shown in orange. The Byes stars of the Paper III My2 and S2 groups, together with their Byes siblings, are circled in a black ellipse in the right panel ($W$ vs $q$ plane); the three Bno stars interpreted as their progenitors are shown in purple. The blue and purple arrows show schematically the evolutions to the S3 and My2/S2 groups of Paper III, respectively.

### *3.11 Family C1*

While it is difficult to identify, for each carbon-rich star of the sample studied in the present work, which family of oxygen-rich stars it comes from, it is relatively straightforward to assign these stars to three different families, which we call C1, C2 and C3, depending on the values taken by their stellar and light curve parameters.

The C1 family collects the carbon-rich stars having amplitudes smaller than 1 mag, 0.4 on average, and colour indices smaller than 2 mag, 1.4 on average, however observed to be never smaller than 1. Whenever measured, their $^{12}C/^{13}C$ isotopic ratio is large, 44 on average, including 3 J-type stars, UV Cam, Y CVn and RY Dra having a low carbon isotopic ratio of 3 to 4 each. They have effective temperatures slightly below 3000 K, ~2770 K on average. However, they have rather low mass-loss rates, ~4 $10^{-7}$ solar masses per year on average, with the possible exception of S Sct, which has been estimated to have a much larger mass-loss rate of 5.6 $10^{-6}$ solar masses per year by Bergeat & Chevallier (2005); it shares with 8 other stars of the family the particularity of having a detached shell. The stars of the C1 family will cool down and either leave the AGB with low amplitudes or start pulsating more clearly with amplitudes exceeding 1.4 mag and enter another carbon-rich family.

### *3.12 Families C2 and C3*

Families C2 and C3 collect carbon-rich stars, which pulsate with amplitudes in excess of 1.4 mag. As illustrated in Figure 11, they cover different regions of effective temperature, above and below 2400 K, defining families C2 and C3, respectively. When the value of the temperature is not available, we use the values of the other stellar and light curve parameters to decide to which family to assign the star. In a few cases, the result is ambiguous, the star might have been assigned to the other family.

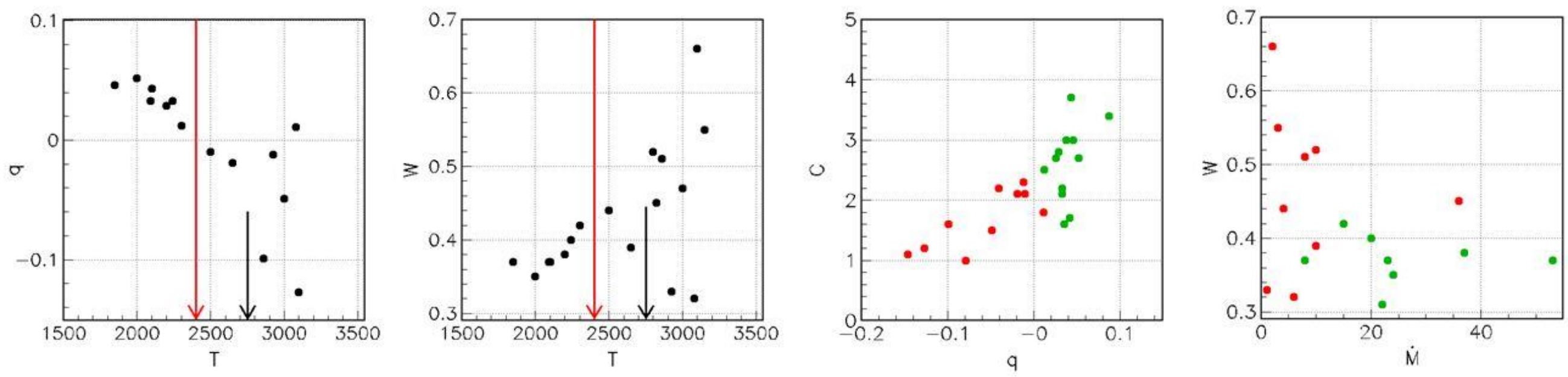

**Figure 11.** The C2 and C3 families. Pair of left panels: distributions of the C2 (on the right of the red arrows) and C3 (on the left of the red arrows) families in the $q$ vs $T$ (left) and $W$ vs $T$ (right) planes; the black arrows show the mean value of the temperature of the C1 family. Pair of right panels: distributions in the $C$ vs $q$ and $\dot{M}$ vs $W$ of the C2 (red) and C3 (green) families.

We build this way two reasonably distinct families; particularly remarkable are the different coverages in $q$, $W$, $C$ and $\dot{M}$, as illustrated in Figure 11. Mean parameters are listed in Table 1: for family C2, $<C>$=1.7 mag, $<A>$=2.5 mag, $<T>$=2860 K, $<\dot{M}>$=0.9 $10^{-6}$ solar masses per year, $<^{12}C/^{13}C>$=17, $<P>$=397 days, $<W>$=0.47, $<q>$=-0.057 and for family C3, $<C>$=2.6 mag, $<A>$=2.9 mag, $<T>$=2120 K, $<\dot{M}>$=3.0 $10^{-6}$ solar masses per year, $<^{12}C/^{13}C>$=50, $<P>$=441 days, $<W>$=0.39, $<q>$=+0.040. All parameters indicate that the C3 family is significantly more evolved than the C2 family, suggesting therefore a likely evolution from family C2 to family C3.

### *3.13 The oxygen-rich to carbon-rich transition*

It is difficult to give with confidence a reliable interpretation of the transition from oxygen-rich to carbon-rich, in part because of the complexity of the mechanisms at play, but also because of our inability to reveal reliably possible pulsations of amplitude smaller than ~0.3 mag: a star pulsating with oscillation amplitude smaller than ~0.2 mag seems to us not to pulsate.

An abundant literature is devoted to the observation and study of carbon-rich stars. Bergeat et al. (2002, 2005) and Tanaka et al. (2007) have shown their location in the HR diagram and recent summaries of current knowledge are given in Marigo et al. (2020), Straniero et al. (2023), Nowotny et al. (2011, 2013), Abia et al. (2020) and Rau et al. (2017). They show that stars of solar metallicity need to have an initial mass between approximately 1.5 and 3.5 solar masses to have a chance to become carbon-rich at the end of their evolution along the TPAGB. Those with lower masses do not experience efficient enough TDU events, if at all, and end their life as M-type or S-type stars (Guandalini & Busso 2006, 2008). Those with larger masses become Hot Bottom Burners with carbon being converted to nitrogen.

However, reliable modelling faces difficulties when attempting to simulate accurately the mechanism of mass loss and the efficiency of TDU events, in particular in relation with the role played by convection. Nowotny et al. (2013) have been able to model the sequence of galactic carbon-rich stars from their blue-end to their red-end and have underscored the important role played by dust formation in the mechanism of mass loss and its effect on the photometric appearance of the star. In particular, Kraemer et al. (2019) have given evidence for significant differences between Mira and semi-regular variables in terms of dust content, whether amorphous carbon, SiC or CS. Of particular relevance to the present work, evidence for a group of carbon-rich stars with small carbon enhancement and mass-loss rate, obtained from a study of the initial to final mass relation (linking the birth mass of a star to the mass of the compact remnant left at its death) by Marigo et al. (2020), has shed some light on the detailed structure of the population of carbon-rich stars in the Galaxy. These stars have spectra dominated by dust-free stellar photospheres, mass-loss rates of a few $10^{-7}$, not exceeding $10^{-6}$ solar masses per year, and expansion velocities not exceeding 10 to 15 km/s. They are mostly classified as semi-regular or irregular, and when a period can reliably be defined it is relatively low, typically 200 days. They have initial masses smaller than ~2 solar masses. When they enter the C/O>1 region, it takes some time for a sufficient quantity of free carbon, not locked in CO molecules to build up, that will make it possible for dust grains to form and significant mass-loss to crop up. Such stars stay therefore much longer in the dust-free, low mass-loss rate region than stars having a larger initial mass.

As long as all carbon-rich stars in our sample are intrinsic, their progenitors can only be from the Ayes or Byes family. Moreover, we expect the progenitor of any carbon-rich star to be at least as warm as it, if not warmer, implying that the Ayes stars that become carbon-rich can only evolve to the C3 family: the mean temperature/rms values of the carbon-rich families are 2770/220 K for C1, 2860/190 K for C2 and 2120/170 K for C3; those of the Ayes and Byes families are 2340/410 K and 2730/340 K, respectively. This also implies that only some Byes stars can transit to the C1 or C2 families. It is then natural to identify part of the C1 stars with those described by Marigo et al. (2020) as low initial mass stars crossing the C/O~1 region, the similarity between their properties being remarkable. While such a scheme is plausible, it leaves some questions unanswered. In particular, the warmer members of the C1 and C2 families can only have warmer members of the Byes family as progenitors but the former significantly outnumber the latter; of course, this may simply be the result of a selection bias of the type discussed by McDonald (2026). More generally, the differences observed between some parameters of the C1 and C2 families do not allow for a clear interpretation of the precise nature of possible transitions from Byes to C1 and/or C2, or from C1 to C2; the mean period of the C2 family is not much larger than that of the Byes family and that of the C1 family is unfortunately unknown.

Another important illustration of the complexity of the oxygen-rich to carbon-rich transition is offered by Marigo (2007), who studies the case of stars having $M_i$ just above the carbon-rich range. She shows that in a narrow range of initial masses, typically 3.5 to 4 solar masses, the effect of TDU events is to weaken or even prevent HBB to occur; namely such stars leave the AGB as oxygen-rich but not as Hot Bottom Burners; this may be the case of some Ayes stars, such as chi Cyg and S Cas.

In summary, while we can state with reasonable confidence that members of the Ayes family which become carbon-rich evolve to the C3 family and that members of the Byes family which become carbon-rich evolve mostly to the C1 and C2 families, and possibly later to the C3 family, details of these transitions remain unclear.

Five stars of the carbon-rich sample are of type J: VX And, WZ Cas, UV Cam, RY Dra and Y CVn. WZ Cas is likely to be a Hot Bottom Burner but the other four are all members of the C1 family. Various scenarios have been proposed for their interpretation, usually invoking some form of extra mixing and generally predicting that such stars are the progeny of low initial mass stars, with $M_i$ well below 2 (Choplin et al. 2024, Abia & Isern 2000, Chen et al. 2007, Hinkle et al. 2016, Abia et al. 2017). However, Y CVn, which is also displaying a detached shell, makes exception and is thought to have lost much of its initial mass, originally in excess of 2 solar masses.

Finally, we note that eleven carbon-rich stars of our sample have detached shells: two C2 stars: ST And and R Scl, and nine C1 stars: UU Aur, AQ And, RT Cap, U Cam, TT Cyg, U Hya, W Ori, TX Psc and Y CVn. Such shells are usually assumed to be the product of recent TPs but Mecina et al. (2014) have shown that a realistic picture of their properties needs to account for a number of complications related to the interaction with both ISM and previous wind. Kastner et al. (2021) have looked for a possible relation between the star bolometric luminosity and the time separating successive thermal pulses; their result suggests that the initial masses of the detached-shell stars are rather on the high side, $M_i$ ~2.5 to 3.5. A surprizing peculiarity is that W Ori is unexpectedly Tc-poor (Cruzalèbes et al. 2013).

## 4. Specific features

The preceding section largely ignored specific features of the stars under study, which are abundantly discussed in the published literature. We review them briefly in the present section.

Binaries are of course common among the stars of the sample under study, and the presence of planetary companions is widespread. An abundant literature is dedicated to the study of their impact on the circumstellar envelope and the wind that it hosts, (Mayer et al. 2013, Danilovich et al. 2025 and references therein); in most cases, the impact on the pulsation and on the light curve is negligible. In some cases, the orbiting companion, when surrounded by a large dust-rich cloud, causes eclipses, which, however, are easily separated from the proper pulses; an abundantly studied example is eta Gem (Torres et al. 2022, Ortiz et al. 2021, Lee et al. 2023), which, together with RR UMi, RW And and T Dra, is an X-ray emitter.

Bow shocks resulting from the interaction between the wind and the ISM are often observed. Cox et al. (2012) have reviewed their properties in a seminal paper and 17 stars of our sample, half of which are members of family C1, are listed by them as producing a bow shock; again, the impact on the pulsation and on the light curve can usually be ignored. In the case of V CVn, the bow shock of the variable dusty stellar wind causes an inverse relationship between the brightness of the light curve and its polarization (Power et al. 2025, Neilson et al. 2014).

Extra mixing, possibly dredging-up Li as already evoked in the context of J-type carbon-rich stars, is suggested in several low initial mass cases such as T Her (Uttenthaler et al. 2011). More generally, the presence and interpretation of s-process elements, including of course technetium, are abundantly discussed in the published literature. In particular recent TP events are commonly invoked to explain changes of the light curve parameters, such as the period (Wood & Zarro 1981, Zijlstra & Bedding 2002). In most cases, the impact on the light curve and stellar parameters, while significant, is not strong enough to invalidate the evolutionary scheme presented in Section 3. Yet, a few cases may be causes of concern. The case of T UMi, which was mentioned in Paper II, is the most spectacular and that which challenges most interpretation (Fadeyev 2022, Molnar et al. 2019, McDonald et al. 2020). Over a period of only two decades, it changed from a regular pulsator with a period of ~300 days and amplitude of ~5 mag to a semiregular dominantly first overtone pulsator with a period of ~200 days and amplitude of ~1.5 mag. Its spectral type decreased from M6 to M4, its colour index from 2.6 to 1.4. Fadeyev (2022) evaluates its initial mass between 1 and 1.5 solar masses, having now decreased by about 30%, using a model that gives a good description of its evolution in terms of the contraction of the envelope in the early phase of a thermal pulse; its being Tc-poor would then imply that the TP is not strong enough to produce significant dredge-up. In the language of the present work, T UMi is a clear member of the Ano family ($W$=0.21, $q$=0.10) before the sudden change and becomes a clear member of the Bno family afterwards. This is particularly puzzling: it is the only explicit and robust example of a transition from the progeny of the A family to the progeny of the B family. A similar, but less spectacular case of sudden change has been observed with RU Vul, which McDonald et al. (2020) have studied in much detail. It differs from T UMi as being metal-poor and having a colour-index that increased rather than decreased, from ~1.6 to ~3.3, the other features being otherwise qualitatively similar but quantitatively less strong and the current mass being evaluated as ~0.6 solar masses. In the language of the present work, RU Vul was a member of the Bno family, with $W$~0.49 and $q$~-0.025, and remained so after the change; however, it evolved against the expected flow.

In a few cases, the light curve displays a characteristic doubly-periodic shape, taking the form of a long period oscillation on top of which an oscillation of much shorter period is superimposed. Such light curves have been identified and abundantly studied in the context of so-called Long Secondary Periods (LSP); Courtney-Barrer et al. (2026) provide a recent review with references to earlier work. The currently favoured interpretation describes the long period oscillations as convective dipoles and the short period oscillations as normal radial pulsations (Saio et al. 2015). To identify such cases requires both a long observation time of several long period cycles and a high density and quality of observations in order to cope with the often-small amplitudes of the short period oscillations. In the sample used in the present work, we reliably identified twelve candidates listed in Table 2. They include one Ayes star (RV Peg), one B1 star (V UMi), two Blow stars (TZ Cyg and OP Her), four Bno stars (S Lep, V465 Cas, U Del and g Her), one C1 star (U Cam) and three C3 stars (S Aur, R Lep and V Hya). A few other stars might have been included, such as X Her, IQ Her, SW Gem and TX Dra, but their cases are less clear. With the exception of RV Peg, these stars are meant to have evolved from the B family if they obey the scheme suggested in the present work. Figure 12 illustrates their distributions in the long vs short amplitude plane, $A_L$ vs $A_S$, and in the long vs short period plane, $P_L$ vs $P_S$. Both periods and amplitudes increase with the degree of evolution. The similarity of the observed features favours a dynamical mechanism over binarity; however, the clear separation in two distinct groups, covering very different ranges in both amplitudes and periods, prevents claiming that they are the results of a same mechanism: it may very well be that the large period cases, from the Ayes and C3 families, are caused by binarity, but such considerations are beyond the scope of the present work. The long periods are typically an order of magnitude larger than the short periods. As these LSP stars span the whole AGB branch, they are clearly not associated with a specific stage of evolution. The central-left panel of Figure 8 shows that the short periods of the LSP candidates are clearly smaller than the general trend displayed by the other Bno stars, suggesting that the presence of a strong dipole attenuates the effect of the radial oscillations. Among the LSP candidates listed in Table 2, V Hya deserves a special mention. As for the other members of family C3, the LSP is usually blamed on binarity in the published literature (Planquart et al. 2024). Knapp et al. (1997, 2000) have presented a detailed study of its case; as mentioned earlier, it is surrounded by a detached shell; it is losing mass at a high rate of several $10^{-6}$ solar masses per year; the circumstellar envelope displays anisotropy varying on a short time scale and hosts a 200 km/s wind; evidence has been obtained for the presence of atomic carbon in the molecular outflow, interpreted as produced by shocks generated by the fast wind; many features suggest that V Hya is currently entering the post-AGB phase.

**Table 2.** LSP candidates. Parameters are defined as in Tables 1 and A1.

| Name/family | Type | $C$ | $A$ | $T$ | $L$ | $R$ | $\dot{M}$ | Tc | $^{12}C/^{13}C$ | $P_S$ | $A_S$ | $P_L$ | $A_L$ |
|---|---|---|---|---|---|---|---|---|---|---|---|---|---|
| S Aur/C3 | C4 | 3.4 | 2.0 | 1940 | 17 | | 45 | | | 610 | 2 | 12200 | 3 |
| U Cam/C1 | C3 | 1.8 | 0.4 | 2750 | 7 | 290 | | | 97 | 220 | 0.5 | 2600 | 0.7 |
| V465 Cas/Bno | M4 | 1.5 | 0.3 | | | | 0.3 | no | | 83 | 0.3 | 867 | 0.7 |
| TZ Cyg/Blow | M6 | 1.9 | 0.4 | | | | | | | 75 | 0.4 | 570 | 0.6 |
| U Del/Bno | M4 | 2.1 | 0.8 | 3160 | 7.1 | 270 | | yes | | 120 | 0.25 | 1200 | 0.7 |
| g Her/Bno | M6 | 0.9 | 0.7 | 3300 | 5.4 | | 1 | no | 13 | 80 | 0.2 | 860 | 0.6 |
| OP Her/Blow | M6S | 0.9 | 0.3 | 3325 | 4.2 | | | yes | 11 | 75 | 0.3 | 720 | 0.4 |
| V Hya/C3 | C6 | 2.7 | 1.5 | 2400 | 14 | | 50 | | 70 | 510 | 2.0 | 6520 | 4 |
| R Lep/C3 | C7 | 2.5 | 2.5 | 2300 | 6 | | 15 | | 34 | 440 | 2.5 | 13900 | 2.5 |
| S Lep/Bno | M6 | 2.2 | 0.8 | | | | 0 | | | 90 | 0.3 | 860 | 0.6 |
| RV Peg/Ayes | M6 | 4.3 | 4.5 | | | | | poss | | 390 | 4.5 | 5500 | 2.5 |
| V UMi/B1 | M5 | 1.0 | 0.5 | | | | | | | 70 | 0.5 | 800 | 0.5 |

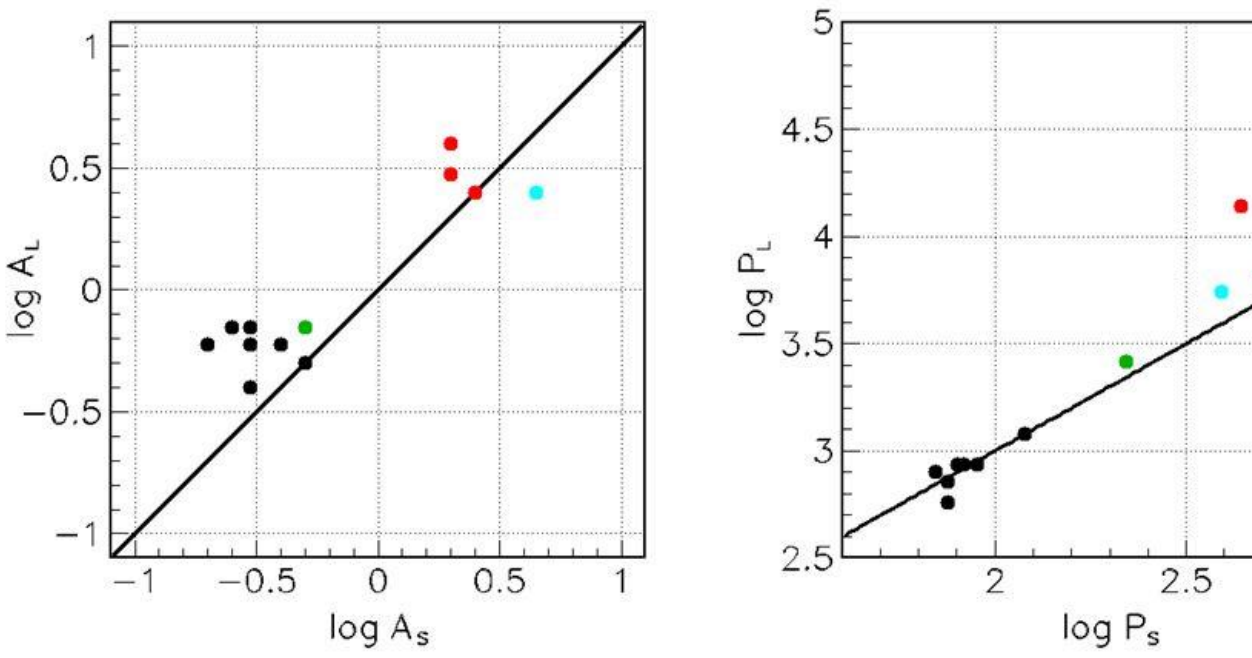

**Figure 12.** Location of the LSP candidates in the $A_L$ vs $A_S$ (left) and $P_L$ vs $P_S$ (right) planes. C3 stars are shown in red, U Cam (C1) in green and RV Peg (Ayes) in cyan. The lines show $A_L = A_S$ and $P_L = 10P_S$ as references.

In several cases, the curve is said to display a double period (Hinkle et al. 2002). In practice, this may correspond to quite different situations, including standard double maximum profiles well described by the ($p$,$q$) parameterisation introduced in Paper III. In the present sample, examples of clear cases are AF Cyg (Bno), AH Dra (Bno), RZ Cyg (Byes) and ST Cam (C1). Some cases are described by a simple power spectrum with two well separated peaks; others display instead contributions of several different frequencies. An analysis of a few typical cases by Kiss et al. (2000) illustrates well the kind of information that can be gathered from an identification of the modes at play. A detailed study is beyond the scope of the present article; we simply illustrate the issue with the light curve of AF Cyg in Figure 13, which has been the subject of a detailed analysis by Ahmad et al. (2025). It displays intervals of reasonably stable pulsations, with periods of ~ 90 or ~180 days, separated by intervals of lower amplitude and less regular variability. It exhibits strong and persistent first overtone activity alongside weaker fundamental mode pulsations, qualitatively well described by the model produced by Ahmad et al. (2025) for a 1 solar mass star having a radius of 167 solar radii, a temperature of 3135 K and a luminosity of 2436 solar luminosities. Indeed, the simulations performed by these authors shed much light on the complexity of the dynamics at play. They use a 3D description including radial and non-radial modes and radial orders up to the second overtone, and accounting for the non-linear interaction between convection and pulsations. They show that the persistence of pulsation nodes appears to depend critically on the interplay between convective and sound velocities, highlighting the sensitivity of the amplitude of pulsations to the stellar structure and convective environment: pulsation amplitudes are influenced by convection-pulsation interactions.

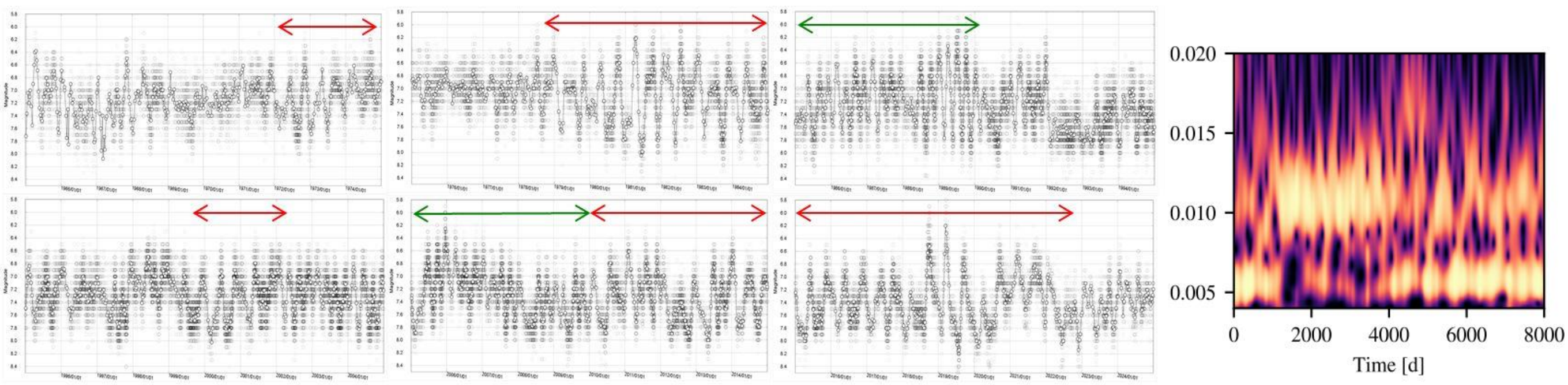

**Figure 13.** Left: light curve of AF Cyg covering six decades from 1965 to 2025 (AAVSO data base); the intervals of relatively stable pulsation are shown with green arrows with a period of ~90 days and with red arrows for a period of ~180 days. Right: Weighted-Wavelet-Z-transform obtained by Ahmad et al. (2025) for a shorter time interval (~22 years).

## 5. Summary and conclusion

In summary, we have considered separately the progenies of stars that enter the AGB already pulsating with amplitudes at the scale of one unit of magnitude and the progeny of stars that enter the AGB with little variability, with amplitudes not exceeding ~0.4 unit of magnitude. Both enter the AGB with early M spectral types and temperatures above 3500 K.

The former, progeny of the A family, leaving aside stars with initial masses larger than some 4 solar masses which become Hot Bottom Burners, includes stars that spend their whole AGB lives as Mira variables and have been studied in Papers I to III. They start climbing the EAGB with their light curves deviating progressively from a sine wave, their temperatures decreasing from above 3500 K to some 3000 K and their periods increasing to some 200 days. They then keep expanding and cooling with increasing periods and decreasing widths of the light pulse. Some may reach temperatures below 2500 K and periods over 500 days without entering the TPAGB, or at least, if they enter it, without experiencing strong TDU events; these keep having a modest mass-loss rate at the scale of $10^{-7}$ solar masses per year and their light curves display narrower and narrower light pulses, with $W$ and $q$ reaching 0.2; they may then leave the AGB as Tc-poor oxygen-rich stars having a low isotopic $^{12}C/^{13}C$ ratio, below ~20. But many will enter the TP AGB at various stages of their expansion and cooling-down phase, becoming Tc-rich stars of the My1 and S1 groups of Paper III with a carbon isotopic ratio in excess of 20, and reaching temperatures at the 2000 K level or below and mass-loss rates well in excess of $10^{-6}$ solar masses per year, possibly becoming cool carbon-rich stars before leaving the AGB.

In contrast, the progeny of the B family maintain a low variability when evolving along the EAGB. They do not start displaying significant pulsations before reaching spectral type M4. Then, some, having spectral types M4 to M6 will start pulsating with amplitudes soon reaching 2 units of magnitudes and typical periods of 200 to 300 days, before entering the TPAGB; they still have low mass-loss rates, at the scale of $10^{-7}$ solar masses per year. The others, which are likely to have low initial masses, below some 1.5 solar masses, will keep expanding and cooling down without experiencing strong TDU events, ultimately leaving the AGB as oxygen-rich stars with low amplitudes not exceeding 0.6 mag. The former will then keep pulsating with increasing amplitudes and periods, eventually entering the TPAGB and becoming Mira variables experiencing strong TDU events, corresponding to groups My2, S2 and S3 of Paper III. Both the Mira variables of the Ayes and Byes families have spectral types close to M6 and display features that give evidence for their having experienced strong TDU events: they are Tc-rich and their $^{12}C/^{13}C$ isotopic ratio has reached ~30; but those of the Byes family differ clearly from those of the Ayes family as having a mean temperature ~400 K higher, a mean period ~100 days shorter, a significantly lower mass-loss rate and remarkably broader light pulses. As a result, those which become carbon-rich enter different families: the carbon-rich progeny of the Ayes family have temperatures below ~2400 K while that of the Byes family have temperatures above it.

The detailed inspection of the light curves of a large number of semi-regular variables, as has been done in the present work, gives evidence for a very broad variety of irregularities; amplitudes may change progressively over many cycles or abruptly over a few; the same applies for periods, however less frequently. In most cases such changes occur at times that succeed each other irregularly. The exercise gives the strong impression of a very complex and disorganized picture, leaving little hope for describing it in simple terms and, a fortiori, for learning about the mechanisms that are driving the pulsations. A very abundant literature is devoted to the study of such light curves, using Fourier and/or wavelet analyses as their main tool; the main outcome is the evidence for complexity: most of the time, using the language of modes of oscillation does not go much beyond renaming the A progeny as fundamental pulsators and revealing important contributions of overtones to the B progeny.

The proposed evolutionary scheme presented in Section 3 is likely to be close to reality: we have given strong arguments in its favour and did not find any against it. Yet, it is very general and gives only a broad picture of the evolution of AGB stars in relation with the properties of their light curves. In particular, the separation in different families of the A and B progenies is not meant to define clearly distinct classes of stars, but simply to show important milestones in the evolution, such as the occurrence of strong TDU events or as the transition from oxygen-rich to carbon-rich.

Our results raise obvious questions related to the separation between the A and B progenies: how strict is it and how reliable is its validity? Particularly puzzling in this context has been the observation of a clear transition from the Ano family to the Bno family by the star T UMi, estimated to have a low initial mass of 1 to 1.5 solar masses and to be in the early phase of a thermal pulse.

Ideally, one would like to assign to the A and B progenies clearly different basic stellar parameters, initial mass and metallicity. Several features suggest that the progeny of the B family have a smaller initial mass than the progeny of the A family. The dependence of luminosity on temperature and period, or the dependence of amplitude on period, illustrated in Figure 14 for stars having both values available show that the A progeny, in addition to being cooler, reach significantly larger luminosities, larger amplitudes and larger periods than the B progeny, consistent with their having larger initial masses. In several instances, we met examples of lower initial mass stars being part of the B progeny and larger initial mass stars being part of the A progeny. But we can mention stars of the A progeny that seem to have a low initial mass, such as W Aql, estimated by De Nutte et al. (2017) to have $M_i$ ~1.6 on the basis of the value of its $^{17}O/^{18}O$ ratio, or stars of the B progeny that seem to have a high initial mass, such as R Hya, estimated by Fadeyev (2023) to have $M_i$~4.5. This prevents claiming with confidence that the A and B progenies are strictly associated with larger and lower initial masses, respectively. It suggests instead that in addition to the value of the initial mass (and most likely initial metallicity) details of the properties of the convective envelope play a role in defining which of the A and B progenies a star belongs to.

Unfortunately, initial masses are not well known. Locating a star in the HR diagram can only be done with significant uncertainties and the tracks predicted by models for stars of different initial masses are close to each other and depend on the details of the model. In a recent thorough review, McDonald (2026) addresses issues related to the evaluation of the ages and initial masses of AGB stars and underscores the difficulties inherent to the exercise.

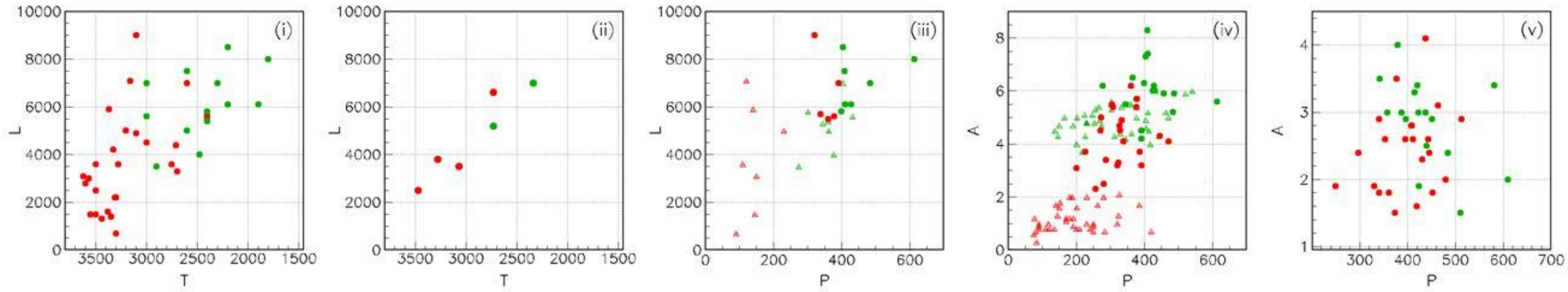

**Figure 14.** Oxygen-rich stars of the A (green) and B (red) progenies. From left to right: Hertzsprung-Russell diagram of i) stars for which the values of both $L$ and $T$ are available, and of ii) the mean values of the different families (in increasing order of $T$, B, Bno, Blow, Ano, Byes and Ayes). Dependence on $P$ of $L$ (iii) and $A$ (iv); full circles are for Ayes and Byes, triangles are for A1, Ano and Bno. Locations (v) of the C2 (red) and C3 (green) families in the $A$ vs $P$ plane.

If it were true that the A and B progenies are simply associated with larger and lower initial masses, respectively, one could sketch the location of the different families in the initial mass vs age plane as illustrated in Figure 15, accounting for the possibility (Marigo 2007) for stars having $M_i$ in the ~3.5 to ~4 range to leave the AGB as oxygen-rich, TDU events preventing the occurrence of HBB. It might then suggest that the A family and its progeny are themselves the progeny of stars that were fuelled by the CNO cycle on the MS while the B family and its progeny would have been fuelled by the pp cycle when they were on the MS; the former entred promptly the Helium Core Burning phase when leaving the MS, while the latter went through the First Dredge Up, the Red Giant Branch and, depending on their metallicity, the Red Clump or the Horizontal Branch. Finally, it would imply that the initial mass sequence of stars leaving the AGB and entering the post-AGB is simply Blow-Bno-Byes-C3-Ayes-Ano-HBB in order of increasing values of $M_i$. But, in the present state of our knowledge, this is pure speculation and is only mentioned here to provide food for thought in inspiring future work. In particular, the outstanding quality and quantity of observations recently produced by Gaia and OGLE, both from the Galaxy and from the Magellanic clouds, can be expected to trigger much progress in our understanding of Long Period Variables. Their study requires a very different approach than used in the present work. They should be able to tell what can be retained of the evolution scheme that we have presented and what must be rejected.

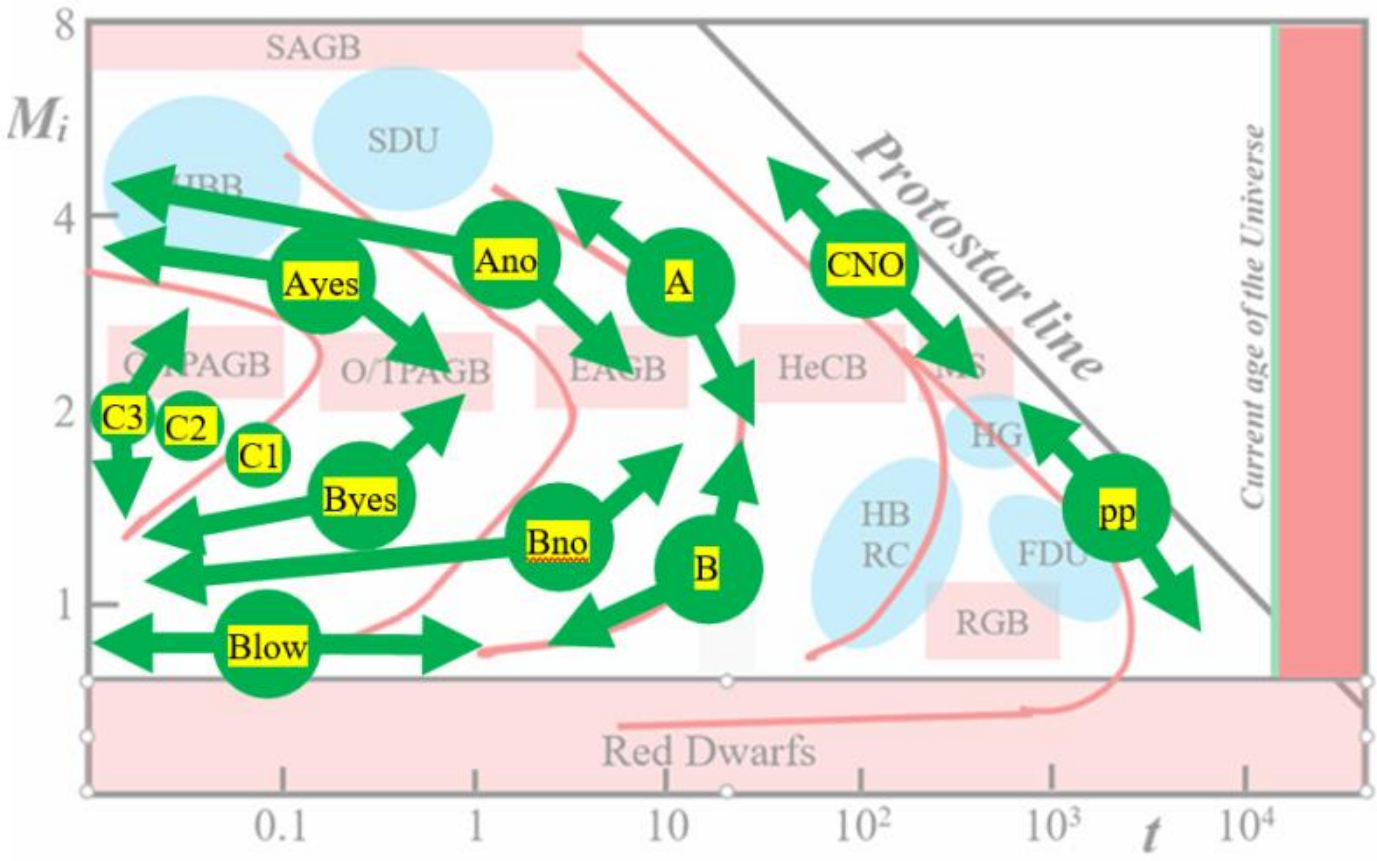

**Figure 15.** A sketch of the suggested location of the star families defined in Section 3 in the $M_i$ vs $t$ plane (Figure 1) under the assumption that the A and B progenies are associated with larger and lower initial masses, respectively.

A global view of the current understanding of the dynamical processes at play is given by the work of Ahmad et al. (2023,2025) and the references quoted by them. They highlight the importance of the non-linearity of the interplay between the self-excited pulsations and the self-consistent convection and the natural emergence of multi-mode pulsations, including both radial and non-radial modes. Much less ambitious, the present work has offered a pragmatic approach matching the limitations of the AAVSO data that prevent defining reliable light curve parameters for amplitudes smaller than 0.2 to 0.3 mag; it has given evidence for two different lines of evolution, one of mostly Mira variables, preferably associated with the larger initial masses and one of mostly semi-regular variables, preferably associated with the lower initial masses; it has clarified a major question that Paper III had left open: the transition from semi-regular variables of the Bno family to Mira variables of the Byes family, thereby confirming the differences between the My1/S1 and My2/S2/S3 groups identified in Paper III; the use of the shape parameters $W$ and $q$ proved to be determinant on this point. Moreover, it has shed light on the main features describing the oxygen-rich to carbon-rich transition, Ayes to C3 and Byes to C1-C2-C3, although several issues still need to be confirmed before the validity of the proposed scheme can be robustly accepted. Some of the results of the present work challenge interpretation and may be a source of inspiration and progress in producing future state-of-the-art simulations.

## Acknowledgements

To the extent that the analyses presented in the present article may have contributed some progress, the credit belongs to the innumerable observers around the world and to the AAVSO, without whom none of the results could have been obtained. We are accordingly deeply indebted to them: to the many observers for the high quality of their observations and to the AAVSO for the outstanding handling and reduction of the data that makes their use particularly easy and efficient. This research has made use of the SIMBAD database, CDS, Strasbourg Astronomical Observatory, France. It is funded by the Vietnam Academy of Science and Technology under grant number VAST 08.02/25-26. Financial support from the World Laboratory and the Rencontres du Vietnam is gratefully acknowledged.

# Appendix

Table A1 lists, for each star and for each family separately, stellar and light curve parameters of the stars studied in the present work. Type is the spectral-type taken from the SIMBAD data-base; *C*, the colour index measured in magnitudes; *A*, the amplitude measured in magnitudes; *T*, the effective temperature measured in Kelvins; *L*, the luminosity measure in $10^3$ solar luminosities; *R*, the stellar radius measured in solar radii; $\dot{M}$, the mass-loss rate measured in $10^{-7}$ solar masses per year; Tc indicates the presence or absence of technetium in the spectrum; $^{12}C/^{13}C$ is the carbon isotopic ratio; 'Ref' gives references to some articles of relevance listed below the table; *P* is the period measured in days; *W* and *q*, the light curve parameters defined in Papers II and III, respectively. When the values are taken from the published literature, they are usually affected by large uncertainties and different authors often quote different values. Accordingly, the values listed in the table are estimates well adapted to the requirements implied by their use in the present work but not to be quoted: the only values to be quoted are from the sources listed in the column 'Ref' and listed below each table.

**Table A1.1** Family A

| Name | Type | *C* | *A* | *T* | *L* | *R* | $\dot{M}$ | Tc | $^{12}C/^{13}C$ | Ref | *P* | *W* | *q* |
|---|---|---|---|---|---|---|---|---|---|---|---|---|---|
| 1 S Aql | M3 | 1.7 | 2.1 | | | | | | | 1 | 147 | 0.47 | -0.034 |
| 2 T Cen | K5-M4 | 1.1 | 2.0 | | | | | | | - | 88 | | |
| 3 RY Cep | M0 | 1.4 | 2.5 | | | | | | | - | 152 | | |
| 4 RY Leo | M2 | 2.4 | 1.8 | | | | | | | 2 | 169 | 0.49 | 0.019 |
| 5 X Lib | M3 | 2.0 | 2.7 | | | | | | | - | 165 | 0.42 | -0.002 |
| 6 RW Sgr | M3 | 2.4 | 1.3 | | | | | no | | - | 190 | | |

1) Cadmus et al., 1991, Astron. J., 101, 104; 2) Kiss et al., 2000, AASS, 145, 283.

**Table A1.2** Family A1

| Name | Type | *C* | *A* | *T* | *L* | *R* | $\dot{M}$ | Tc | $^{12}C/^{13}C$ | Ref | *P* | *W* | *q* |
|---|---|---|---|---|---|---|---|---|---|---|---|---|---|
| 7 RS Aqr | M2 | 1.5 | 3.7 | | | | | | | - | 217 | 0.39 | -0.036 |
| 8 T Aqr | M2 | 1.4 | 5.0 | | | | | no | | - | 202 | 0.34 | -0.049 |
| 9 R Boo | M4 | 1.7 | 5.1 | | | | | no | | - | 224 | 0.31 | -0.018 |
| 10 T Col | M4 | 1.7 | 4.3 | | | | | no | | - | 226 | 0.33 | -0.028 |
| 11 RT Cyg | M2 | 1.5 | 4.7 | | | 227 | | dbfl | | 1 | 190 | 0.38 | -0.057 |
| 12 T Her | M2.5 | 1.5 | 5.0 | 3400 | | | | no | | 2 | 165 | 0.34 | -0.046 |
| 13 RY Oph | M3 | 1.7 | 4.7 | | | | | no | | - | 150 | 0.32 | -0.021 |
| 14 X Peg | M2.5 | 1.3 | 4.0 | | | | | no | | - | 201 | 0.39 | -0.085 |
| 15 R Vir | M3.5 | 1.8 | 4.3 | 3500 | | 130 | 1.0 | - | | 3,4,5 | 146 | 0.37 | -0.065 |
| 16 R Vul | M3 | 1.8 | 4.5 | | | | | no | | - | 137 | 0.37 | -0.061 |

1) Baylis-Aguirre et al., 2025, MNRAS, 544, 2048; 2) Uttenthaler et al., 2011, A&A, 531, A88; 3) Kervella et al., 2014, A&A, 564A, 88K; 4) Shetye et al., 2025, A&A, 702, A39; 5) Cheng et al., 2025, MNRAS, 538, 3144.

**Table A1.3** Family HBB

| Name | Type | *C* | *A* | *T* | *L* | *R* | $\dot{M}$ | Tc | $^{12}C/^{13}C$ | Ref | *P* | *W* | *q* |
|---|---|---|---|---|---|---|---|---|---|---|---|---|---|
| 17 SV Cas | M5 | 2.7 | 2.2 | | | | 2.4 | no | 10 | 1,2 | 450 | 0.35 | |
| 18 WZ Cas | C9,2 | 1.0 | 0.6 | 3300 | | | 1.3 | | 5 | 3,4,5,6,7,8,9 | 366 | | |
| 19 R Cen | M4 | 2.1 | 5.2 | 3000 | | | | no | | 1,2 | 548 | 0.44 | -0.169 |
| 20 DH Cyg | M6 | 2.7 | 2.8 | | | | | - | | 2 | 531 | 0.51 | |
| 21 RU Cyg | M6/7 | 2.5 | 1.0 | 3150 | 78 | | | Yes/no | | 1,2,9,10 | 468 | 0.58 | |

| 22 R Nor | M3 | 2.3 | 5.6 | 3200 | | | | no | | 2,11 | 498 | 0.35 | -0.137 |
|---|---|---|---|---|---|---|---|---|---|---|---|---|---|
| 23 U CMi | M4 | 2.0 | 4.0 | | | | | no | | - | 409 | 0.47 | -0.164 |

1) Garcia-Hernandez et al., 2013, A&A, 555, L3; 2) Nhung et al., 2025, ApJ, 994, 130; 3) Tanaka et al., 2007, PASJ, 59, 939; 4) Bergeat. & Chevallier, 2005, A&A, 429, 235; 5) Herbig, 1955, PASP, 67, 181; 6) Keenan & Bidelman, 1979, PASP, 91, 365; 7) Lebzelter et al., 2005, A&A, 540, 295; 8) Harris et. al., 2006, MNRAS, 367, 400; 9) Cadmus, 2021, Astron. J., 161, 75; 10) Lebzelter & Hron, 1999, AA, 351, 533; 11) Uttenthaler et al., 2011, A&A, 531, A88.

**Table A1.4** Family Ano

| Name | Type | *C* | *A* | *T* | *L* | *R* | $\dot{M}$ | Tc | $^{12}C/^{13}C$ | Ref | *P* | *W* | *q* |
|---|---|---|---|---|---|---|---|---|---|---|---|---|---|
| 24 R Aql | M6.5 | 2.3 | 4.7 | 2900 | 3.5 | | 11 | no | 8 | 1,2,3,4 | 274 | 0.31 | -0.012 |
| 25 RU Aur | M9 | 3.6 | 5.0 | | | | | | | - | 470 | 0.21 | 0.192 |
| 26 V Cam | M7 | 3.6 | 5.9 | | | | | | | 5,6 | 521 | 0.17 | 0.190 |
| 27 R Cas | M6.5 | 2.1 | 6.0 | 3000 | 5.6 | | 9 | no | 12 | 7,8 | 432 | 0.21 | 0.094 |
| 28 V Cas | M5 | 1.6 | 4.8 | | | | | no | | 9,10 | 229 | 0.28 | 0.014 |
| 29 Y Cas | M7.5 | 3.0 | 4.5 | | | | | | | 11 | 413 | 0.26 | 0.105 |
| 30 X Cep | M4 | 3.2 | 6.0 | | | | | | | 6 | 540 | 0.16 | 0.217 |
| 31 R Cnc | M6.5 | 1.9 | 4.4 | 2600 | 5.0 | | 0.2 | no | 17 | 12,13,14,15 | 362 | 0.25 | 0.061 |
| 32 S CrB | M6.5 | 2.8 | 5.5 | 2400 | 5.4 | | 2.3 | poss | | 16 | 360 | 0.26 | 0.081 |
| 33 R CVn | M6.5 | 2.1 | 4.4 | | | | | no | | 9 | 329 | 0.30 | -0.002 |
| 34 SX Cyg | M7 | 2.3 | 5.2 | | | | | | | - | 410 | 0.20 | 0.156 |
| 35 R Dra | M4.5 | 2.1 | 5.1 | | | | | dbfl | | - | 247 | 0.30 | 0.017 |
| 36 T Eri | M5 | 1.9 | 4.8 | | | | | no | | - | 253 | 0.31 | -0.027 |
| 37 W Eri | M7 | 2.4 | 5.4 | 2950 | | | | no | | 17 | 376 | 0.22 | 0.118 |
| 38 U Her | M6.5 | 2.5 | 4.9 | 3000 | 7.0 | 320 | 4 | no | 19 | 18,19,4,20 | 404 | 0.21 | 0.121 |
| 39 RU Hya | M6.5 | 3.3 | 5.4 | | | | | no | | 21,11 | 332 | 0.24 | 0.101 |
| 40 R Leo | M7 | 1.2 | 4.3 | 2700 | | 340 | 1 | no | 10 | 22,23,7,24 | 312 | 0.26 | 0.065 |
| 41 R Oct | M5.5 | 2.0 | 4.6 | | | | | | | 25 | 405 | 0.18 | 0.134 |
| 42 U Oct | M6 | 1.8 | 5.3 | | | | | no | | - | 303 | 0.27 | 0.016 |
| 43 R Oph | M5 | 2.1 | 5.6 | | | | | no | | 26 | 303 | 0.27 | 0.011 |
| 44 R Peg | M6 | 2.3 | 5.0 | 2480 | 4.0 | 350 | | no | 10 | 27 | 377 | 0.24 | 0.072 |
| 45 W Peg | M6.5 | 2.0 | 4.2 | | 5.3 | | 4.5 | no | 17 | 28 | 345 | 0.28 | 0.013 |
| 46 R Sgr | M4 | 1.5 | 4.9 | | | | | no | | - | 269 | 0.31 | -0.015 |
| 47R Tel | M5 | 2.5 | 5.2 | | | | | | | - | 462 | 0.16 | 0.131 |
| 48 R Tri | M3.5 | 1.7 | 5.4 | 2400 | | | 1.1 | no | | 29 | 266 | 0.25 | 0.008 |
| 49 R UMa | M5 | 2.9 | 5.5 | | 5.8 | | 1.3 | | | 30,31,28 | 301 | 0.27 | 0.073 |
| 50 RR UMa | M4 | 1.6 | 4.8 | | | | | no | | - | 231 | 0.29 | 0.053 |
| 51 T UMa | M4 | 2.4 | 5.3 | | | | | no | | - | 256 | 0.31 | 0.023 |
| 52 Y Vel | M9 | 2.4 | 4.3 | | | | | | | - | 442 | 0.17 | 0.227 |
| 53 Z Vel | M9 | 2.1 | 4.0 | | | | | | | - | 417 | 0.25 | 0.047 |
| 54 RS Vir | M8 | 3.2 | 5.5 | 2900 | | | | no | | 17 | 354 | 0.28 | 0.066 |

1) Baudry et al., 2023, A&A, 674, 125; 2) Zhao-Geisler et al., 2012, A&A, 545, A56; 3) Malfait et al., 2024, A&A, 691, A57; 4) Marinho et al., 2024, A&A, 688, A143; 5) Wiesemeyer et al., 2009, A&A, 498, 3; 6) Cho & Kim, 2012, Astron. J., 144, 129; 7) Brand et al., 2025, A&A, 699, A326; 8) Danilovich et al., 2016, A&A, 588, A119; 9) Shetye et al., 2025, A&A, 702, A39; 10) Guerrero & Ortiz, 2020, MNRAS, 491, 680; 11) Baek et al., 2025, AJ, 170, 84; 12) Cotton et al., 2009, APJSS, 185, 585; 13) Wittkowski et al., 2016, A&A, 587, A12; 14) Karovikova et al., 2013, A&A, 560, A75; 15) Wittkowski et al., 2011, A&A, 532, L7; 16) Danilovich et al., 2015, A&A, 581, A60; 17) Uttenthaler & Lebzelter, 2010, A&A, 510, A62; 18) Winnberg et al., 2024, A&A, 686, A251; 19) Danilovich et al., 2025, A&A, 704, A341; 20) Wallström et al., 2024, A&A, 681, 50; 21) Baudry et al., 2018, A&A, 609, A25; 22) Vlemmings et al., 2024, A&A, 686, A274; 23) Gorai et al., 2025, A&A, 704, A31; 24) Paladini et al., 2017, A&A, 600, A136; 25) Price et al., 2010, ApJSS, 190, 203 ; 26) Kim, et al., 2010, ApJSS, 188, 209 ; 27) Wittkowski et al., 2018, A&A, 613, L7; 28) Alonso-Hernández et al., 2024, A&A, 684, A77; 29) Baylis-Aguirre et al., 2025, MNRAS, 544, 2048; 30) Ramstedt et al., 2012, A&A, 543, A147; 31) Ortiz et al., 2019, MNRAS, 482, 4697.

**Table A1.5** Family Ayes

| Name | Type | *C* | *A* | *T* | *L* | *R* | $\dot{M}$ | Tc | $^{12}C/^{13}C$ | Ref | *P* | *W* | *q* |
|---|---|---|---|---|---|---|---|---|---|---|---|---|---|
| 55 R And | S5 | 2.6 | 7.4 | 1900 | 6.1 | 800 | 9 | yes | 33 | 1,2,3,4 | 410 | 0.22 | 0.045 |
| 56 RW And | M5 | 2.4 | 6.0 | - | | | | | | 5 | 430 | 0.20 | 0.157 |
| 57 W And | S7 | 1.8 | 6.3 | 2400 | 5.8 | 800 | 2 | yes | | 1,3 | 398 | 0.22 | 0.079 |
| 58 X And | S2.5 | 2.2 | 5.5 | - | | | | | | - | 345 | 0.27 | 0.039 |
| 59 W Aql | S6 | 3.2 | 5.2 | 2300 | 7.0 | 500 | 30 | yes | 26 | 6,7,8 | 483 | 0.26 | 0.057 |
| 60 R Aur | M6.5 | 2.1 | 5.9 | 2400 | | | | yes | 33 | 9 | 457 | 0.21 | 0 |
| 61 S Cas | S3 | 3.6 | 5.6 | 1800 | 8.0 | | 30 | | 32 | 1,3,10 | 613 | 0.20 | 0.088 |
| 62 U Cas | S4.5 | 1.2 | 6.2 | - | | | | | | - | 277 | 0.25 | 0.028 |
| 63 Chi Cyg | S6 | 2.1 | 8.3 | 2600 | 7.5 | 410 | 8 | yes | 36 | 11,12,13 | 408 | 0.18 | 0.019 |
| 64 R Cyg | S4 | 2.9 | 6.2 | 2200 | 6.1 | 540 | 9 | yes | 34 | 3,4 | 427 | 0.24 | 0.036 |
| 65 RU Her | M6 | 2.3 | 5.9 | - | | | 16 | yes | 25 | 14,15 | 486 | 0.18 | 0.122 |
| 66 R Hor | M5 | 2.1 | 7.3 | 2200 | 8.5 | 300 | 6 | yes | | 1,16 | 404 | 0.23 | 0.106 |
| 67 RV Peg | M6 | 4.3 | 4.5 | | | | | poss | | 9 | 390 | 0.25 | 0.059 |
| 68 S Pic | M7 | 2.9 | 6.0 | | | | | | | - | 425 | 0.21 | 0.128 |
| 69 R Ser | M5 | 2.0 | 6.5 | | | | 2 | yes | 14 | 15 | 365 | 0.24 | 0.061 |
| 70 T Sgr | S4.5 | 1.5 | 4.2 | - | | | 2 | yes | | 4 | 391 | 0.27 | 0.010 |
| 71 S Vir | M6 | 2.0 | 5.7 | 3280 | | | | yes | | - | 377 | 0.24 | 0.055 |

1) Danilovich et al., 2015, A&A, 581, A60; 2) Danilovich et al., 2018, A&A, 617, A132; 3) Ramstedt et al., 2009, A&A, 499, 515; 4) Wallerstein et al., 2011, A&A, 535A, 101; 5) Ortiz & Guerrero, 2021, ApJ, 912, 9; 6) Danilovich et al., 2024, Nature Astronomy, 8, 308; 7) Danilovich et al., 2025, A&A, 704, A341; 8) Wallström et al., 2024, A&A, 681, 50; 9) Cho & Kim, 2012, Astron. J., 144, 129; 10) Son et al., 2025, Astron. J., 170, 266; 11) Hoai et al., 2024, A&A, 692, A86; 12) Schöier et al., 2011, A&A, 530, A83; 13) Lacour et al., 2009, ApJ, 707, 632; 14) Alonso-Hernández et al., 2024, A&A, 684, A77; 15) Hinkle et al., 2016, ApJ, 825, 38H; 16) Ramstedt et al., 2020, A&A, 640, 133.

**Table A1.6** Family B

| Name | Type | *C* | *A* | *T* | *L* | *R* | $\dot{M}$ | Tc | $^{12}C/^{13}C$ | Ref | *P* | *W* | *q* |
|---|---|---|---|---|---|---|---|---|---|---|---|---|---|
| 72 W Boo | M3 | 0.4 | 0.4 | 3600 | 2.8 | 129 | | no | | 1 | | | |
| 73 VZ Cam | M4 | 0.2 | 0.2 | 3550 | 1.5 | 89 | | - | 14 | 2,3,4 | | | |
| 74 V393 Cas | M0 | 0.4 | 0.2 | | | | | no | | - | | | |
| 75 DM Cep | M3 | 0.3 | 0.2 | | | | | - | | - | | | |
| 76 RR CrB | M3 | 1.0 | 0.3 | 3310 | 2.2 | | | no | | 5,6,7 | | | |
| 77 V1339 Cyg | M4 | 0.3 | 0.2 | | | | | no | | - | | | |
| 78 BQ Gem | M4 | 0.2 | 0.3 | | 2.4 | | 0.8 | no | | 8 | | | |
| 79 Eta Gem | M2 | 0.0 | 0.2 | 3500 | 3.6 | 157 | | no | | 9,10,11,12,13,14 | | | |
| 80 V566 Her | M0 | 0.3 | 0.3 | | | | | no | | 15 | | | |
| 81 R Lyr | M4.5 | 0.5 | 0.2 | 3300 | | | | no | 6.4 | 16,17,18,19,20 | | | |
| 82 Rho Per | M4 | 0.4 | 0.2 | 3570 | 3.0 | | 0.1 | no | 9.7 | 16, 21,22,23 | | | |
| 83 TV Psc | M3 | 0.7 | 0.3 | 3500 | 2.5 | | 0.1 | no | | 24 | | | |
| 84 V380 Sco | M2 | 0.4 | 0.6 | | | | | no | | - | | | |
| 85 ST UMa | M4 | 1.3 | 0.5 | | 2.8 | | | no | | 25,26,5,27 | | | |
| 86 RR UMi | M4.5 | 0.3 | 0.3 | 3440 | 1.3 | 114 | | no | 13 | 10,15,28,5, 29,16,27 | | | |

1) Baines et al., 2018, AJ, 155, 30; 2) Lebzelter et al., 2019, ApJ, 886, 1; 3) Brown et al., 2018, A&A, 616, A1; 4) Tabur et al., 2009, MNRAS, 400, 1945; 5) Baek et al., 2025, AJ, 170, 84; 6) Ueta et al., 2019, PASJ, 71, 4U; 7) McDonald et al., 2012, MNRAS, 427, 343; 8) McDonald et al., 2018, MNRAS, 481, 4984; 9) Torres & Sakano, 2022, MNRAS, 516, 2514; 10) Ortiz & Guerrero, 2021, ApJ, 912, 9; 11) Lee et al., 2023, A&A, 678, 106; 12) Ruban & Arkharov, 2010, ApJ, 53, 523; 13) Richichi & Calamai, 2003, A&A, 399, 275; 14) Hünsch et al., 1998, AA, 330, 225; 15) Famaey et al., 2009, A&A, 498, 627; 16) Tsuji, 2008, A&A, 489, 1271; 17) Sharma et al., 2016, A&A, 585, 64; 18) Harper et al., 2013, MNRAS, 428, 2064; 19) Sloan et al., 2015, ApJ, 811, 45; 20) Rayner et al., 2009, ApJSS, 185, 289; 21) Mennesson et al., 2011, ApJ, 743, 178; 22) Chou et al., 2007, ApJ, 670, 346; 23) Chou et al., 2010, ApJ, 708, 1290 ; 24) Rastau et al., 2023, A&A, 680, A12; 25) Ortiz et al., 2019, MNRAS, 482, 4697 ; 26) Glass & Van Leeuwen, 2007, MNRAS, 378, 1543; 27) Alonso-Hernández et al., 2024, A&A, 684, A77; 28) Lebzelter et al., 2019, ApJ, 886, 117; 29) Diaz-Luis et al., 2019, A&A, 629, A9.

**Table A1.7** Family B1

| Name | Type | *C* | *A* | *T* | *L* | *R* | $\dot{M}$ | Tc | $^{12}C/^{13}C$ | Ref | P | W | q |
|---|---|---|---|---|---|---|---|---|---|---|---|---|---|
| 87 V450 Aql | M5 | 0.2 | 0.3 | 3300 | 2.2 | | | no | 14 | - | | | |
| 88 RV Boo | M5 | 2.0 | 0.4 | | | | | no | | 1 | | | |
| 89 SS Cep | M5 | 1.2 | 0.3 | | 5.1 | | 6 | - | | 2,3,4 | | | |
| 90 RT Cnc | M5 | 1.5 | 0.5 | | 2.2 | | 0.3 | - | | 2,3,5,6 | | | |
| 91 TU CVn | M5 | 0.3 | 0.2 | | | | | no | | 6,7 | | | |
| 92 CZ Del | M5 | 1.0 | 0.3 | | | | | no | | - | | | |
| 93 RR Eri | M5 | 1.0 | 0.4 | | 4.0 | | 4.1 | no | | 1,5 | | | |
| 94 Z Eri | M5 | 1.3 | 0.4 | | | | 0 | no | | 4,6,8 | | | |
| 95 SW Gem | M5 | 1.5 | 0.6 | | | | | | | - | | | |
| 96 UW Her | M5 | 1.3 | 0.4 | | | | | - | | - | | | |
| 97 V Hor | M5 | 0.6 | 0.5 | | | | | - | | - | | | |
| 98 SV Lyn | M5 | 1.5 | 0.4 | | | | | - | | - | | | |
| 99 TT Per | M5 | 1.3 | 0.4 | | | | | | | - | | | |
| 100 Tau4 Ser | M5 | 1.4 | 0.3 | | | | | - | | 9 | | | |
| 101 TX Tau | M5 | 0.7 | 0.6 | | | | | | | - | | | |
| 102 V UMi | M5 | 1.0 | 0.5 | | | | | - | | - | 70 | | |

1) Shetye et al., 2025, A&A, 702, A39; 2) Ortiz et al., 2019, MNRAS, 482, 4697; 3) Olofsson et al., 2002, A&A, 391, 1053; 4) Diaz-Luis et al., 2019, A&A, 629, A94; 5) Alonso-Hernández et al., 2024, A&A, 684, A77; 6) Baek et al., 2025, AJ, 170, 84; 7) Glass & Van Leeuwen, 2007, MNRAS, 378, 1543; 8) Tabur et al., 2009, MNRAS, 400, 1945; 9) Smith, 2025, ApJ, 170, 332.

**Table A1.8** Family Blow

| Name | Type | *C* | *A* | *T* | *L* | *R* | $\dot{M}$ | Tc | $^{12}C/^{13}C$ | Ref | *P* | *W* | *q* |
|---|---|---|---|---|---|---|---|---|---|---|---|---|---|
| 103 Theta Aps | M6.5 | 0.5 | 0.6 | 2700 | 3.3 | 208 | 1 | no | | 1,2,3,4,5 | 120 | | |
| 104 RZ Ari | M6 | 0.5 | 0.3 | 3350 | 1.4 | | | no | 7 | 6,7,8,9,10,11,12 | | | |
| 105 RX Boo | M7.5 | 1.6 | 0.4 | 2750 | 3.6 | 270 | 6 | no | 16 | 13,14,15 | | | |
| 106 AA Cas | M6 | 0.7 | 0.3 | 3260 | | | | - | | 16 | | | |
| 107 RS CrB | M7 | 2.2 | 0.7 | | | | | | | 17,18 | 330 | | |
| 108 TZ Cyg | M6 | 1.9 | 0.4 | | | | | | | - | 75 | | |
| 109 CT Del | M7 | 1.2 | 0.4 | | | | | - | | - | | | |
| 110 EU Del | M6 | 0.7 | 0.4 | 3380 | 1.6 | 128 | 0.5 | no | 14 | 19,20 | | | |
| 111 R Dor | M8 | 1.5 | 0.8 | 2710 | 4.4 | 298 | 2 | - | 10 | 21.22 | | | |
| 112 S Dra | M6 | 1.8 | 0.5 | | | | 1 | | | 23 | 180 | | |
| 113 OP Her | M6S | 0.9 | 0.3 | 3325 | 4.2 | | | yes | 11 | 8,24,25,26,27 | 75 | | |
| 114 ST Her | M6S | 2.0 | 0.6 | | | | 2 | yes | 25 | 28,29,30 | 175 | | |
| 115RX Lep | M6 | 1.4 | 0.5 | | | | 2 | no | | 31,32,33 | | | |
| 116 T Mic | M7 | 1.6 | 0.5 | 3100 | 4.9 | | 1 | - | | 34,35,36,37,38,39 | 350 | | |
| 117 Y Uma | M7 | 1.6 | 0.5 | 3100 | | | 3 | - | | 32 | | | |
| 118 SW Vir | M7 | 1.2 | 0.5 | 3000 | 4.5 | 244 | 4 | yes | 26 | 8,11,21,33,40, 41,42,43 | 180 | | |

1) Ohnaka, 2014, A&A, 561, 47; 2) Moon et al., 2008, JAAVSO, 36/1, 77; 3) Cox et al., 2012, A&A, 537, 35; 4) Maercker et al., 2022, A&A, 663, A64; 5) Mayer et al., 2013, A&A, 549, 69; 6) Konstantinova-Antova et al., 2010, A&A, 524, 57; 7) Konstantinova-Antova et al., 2024, A&A, 681, 36; 8) Lebzelter et al., 2019, ApJ, 886, 117; 9) Maas et al., 2016, Astron. J, 152, 196; 10) Maas & Pilachowski, 2018, Astron. J, 156, 2; 11) Tsuji, 2008, A&A, 489, 1271; 12) Rayner et al., 2009, ApJSS, 185, 289; 13) Kamezaki et al., 2012, PASJ, 64/1, 7; 14) Dillon & Speck, 2022, Bull. AAS, 54/6, 4; 15) Gómez-Garrido et al., 2020, A&A, 642, A213; 16) Ghosh et al., 2019, MNRAS, 484, 4619; 17) Trabucchi et al., 2023, A&A, 680, A36; 18) Mennesson et al., 2005, ApJ, 634, 169; 19) McDonald & Zijlstra, 2016, ApJL, 823, L38; 20) Pavlenko et al., 2020, A&A, 633, 52; 21) Ohnaka et al., 2019, A&A, 621, A6; 22) Khouri et al., 2024, A&A, 685, 11; 23) Diaz-Luis et al., 2019, A&A, 629, A94; 24) Shetye et al., 2025, A&A, 702, A39; 25) Halabi & El Eid, 2015, MNRAS, 451, 2957; 26) Jorissen et

al., 1993, AA, 271, 463; 27) Percy et al., 2009, JAAVSO, 37, 71; 28) Hony et al., 2009, A&A, 501, 609; 29) Wallerstein et al., 2011, A&A, 535A, 101; 30) Adelman & Dennis, 2005, Balt. A., 14, 41; 31 Libert et al., 2008, A&A, 491, 789; 32) Matthews et al., 2013, AJ, 145, 97; 33) Baek et al., 2025, AJ, 170, 84; 34) Paladini et al., 2017, A&A, 600, A136; 35) Baudry et al., 2023, A&A, 674, 125; 36) Danilovich et al., 2025, A&A, 704, A341; 37) Gottlieb et al., 2022, A&A, 660, 94; 38) Wallström et al., 2024, A&A, 681, 50; 39) Montargès et al., 2023, A&A, 671, A96; 40) Khouri et al., 2020, A&A, 635, A200; 41) Hadjara et al., 2019, MNRAS, 489, 2595; 42) Richichi et al., 2020, MNRAS, 498, 2263R; 43) Famaey et al., 2009, A&A, 498, 627.

**Table A1.9** Family Bno

| Name | Type | *C* | *A* | *T* | *L* | *R* | $\dot{M}$ | Tc | $^{12}C/^{13}C$ | Ref | *P* | *W* | *q* |
|---|---|---|---|---|---|---|---|---|---|---|---|---|---|
| 119 RU And | M5.5 | 1.3 | 2.0 | | | | | - | | 1 | 230 | | |
| 120 RV And | M4 | 1.3 | 1.1 | | | | | | | 2 | 170 | 0.50 | 0 |
| 121 TV And | M4 | 2.3 | 1.0 | | | | | | | - | 118 | | |
| 122 T Ari | M6 | 1.7 | 2.1 | | | | 0.2 | no | | 3 | 326 | 0.50 | -0.074 |
| 123 RS Aur | M5.7 | 2.7 | 1.2 | | | | | - | | - | 170 | 0.49 | 0.005 |
| 124 RW Boo | M5 | 1.4 | 0.7 | | | | 2.2 | - | | 4,5,6,7 | 420 | | |
| 125 U Boo | M4.5 | 1.6 | 1.6 | | | | | | | 2,8 | 201 | 0.45 | -0.023 |
| 126 RR Cam | M5 | 1.4 | 0.8 | | | | | | | 9 | 210 | | |
| 127 RS Cam | M4 | 1.6 | 0.9 | 3300 | 0.7 | | | | | 10 | 90 | | |
| 128 RY Cam | M3 | 0.5 | 0.8 | | | | | | | - | 135 | 0.53 | 0.004 |
| 129 V465 Cas | M4 | 1.5 | 0.3 | | | | 0.3 | no | | 11,12 | 83 | | |
| 130 RS Cnc | M6S | 1.1 | 1.0 | 3200 | 5.0 | 225 | 1.2 | yes | | 13,14 | 230 | | |
| 131 V CVn | M4 | 3.1 | 1.2 | | | | | | | 15,16,17,18,19 | 191 | 0.54 | -0.056 |
| 132 AF Cyg | M5 | 1.7 | 1.0 | | | | | no | | 20,21,22,23 | 90 | | |
| 133 W Cyg | M4 | 1.5 | 1.7 | 3370 | 5.9 | 227 | 0.3 | yes | 15 | 19; 24,25,26 | 140 | | |
| 134 U Del | M4 | 2.1 | 0.8 | 3160 | 7.1 | 270 | | yes | | 27,28,29,30,31, 32,33,34,35 | 120 | | |
| 135 AH Dra | M5 | 1.4 | 0.9 | | | | | no | | 11 | 191 | 0.53 | 0.041 |
| 136 TT Dra | M5 | 1.3 | 1.2 | | | | | - | | - | 180 | | |
| 137 TX Dra | M5 | 1.5 | 1.2 | | | | | no | | 36 | 77 | | |
| 138 g Her | M6 | 0.9 | 0.7 | 3300 | 5.4 | | 1.0 | no | 13 | 37,38,39 | 80 | | |
| 139 IQ Her | M3 | 1.8 | 0.6 | | | | 0.1 | no | | 7 | 75 | | |
| 140 RR Her | SC5.5 | 1.3 | 0.8 | 3450 | | | | | 43 | 40,41 | 240 | | |
| 141 X Her | M6 | 1.7 | 0.8 | 3280 | 3.6 | 183 | 1.4 | no | | 7,42,43,44,45,46,47 | 110 | | |
| 142 RT Hya | M6 | 1.6 | 1.0 | | | | | - | | 7,11 | 250 | | |
| 143 S Lep | M6 | 2.2 | 0.8 | | | | 0 | - | | 7,36 | 90 | | |
| 144 U LMi | M6 | 1.5 | 2.0 | | | | | - | | - | 280 | | |
| 145 X Mon | M4 | 1.7 | 1.8 | | | | | poss | | 48 | 152 | 0.45 | 0.005 |
| 146 BQ Ori | M5 | 1.4 | 0.9 | 3145 | | | | - | | 49 | 250 | | |
| 147 S Pav | M8 | 1.5 | 1.7 | 3100 | | | 1.0 | no | | 31,32,33, 50,51 | 385 | 0.57 | -0.065 |
| 148 RU Per | M4 | 2.2 | 1.0 | | | | | | | - | 320 | | |
| 149 $L_2$ Pup | M5 | 1.2 | 1.3 | 3500 | 1.5 | 123 | 0.8 | - | | 52,53,54 | 145 | | |
| 150 W Tau | M3 | 1.6 | 1.7 | | | | | | | - | 262 | 0.44 | 0.063 |
| 151 RX UMa | M5 | 1.2 | 2.0 | | | | | - | | - | 180 | | |
| 152 RY UMa | M3 | 2.2 | 0.7 | 3330 | | | | no | | 9 | 283 | | |
| 153 RZ UMa | M5 | 1.8 | 0.7 | | | | 1.2 | | | 4 | 250 | | |
| 154 V UMa | M5 | 1.5 | 0.8 | | | | | | | - | 200 | | |
| 155 Z UMa | M5 | 1.4 | 2.0 | | | | ~0 | no | | 7,11,18 | 188 | 0.43 | 0.028 |
| 156 R UMi | M7 | 1.6 | 1.3 | 2875 | | | | - | | 11,16,55 | 324 | 0.53 | -0.036 |
| 157 RU Vul | M4 | ~1.6 | 1.6 | 3620 | 3.1 | 142 | | no | | 56,57,58 | 150 | 0.49 | -0.025 |

1) Cadmus, 2021, Astron. J., 161, 75; 2) Cadmus et al., 1991, AJ, 101, 104; 3) Uttenthaler, 2024, A&A, 692, A224; 4) Alonso-Hernández et al., 2024, A&A, 684, A77; 5) Sahai & Stenger, 2023, Astron. J., 165, 229; 6) Guerrero & Ortiz, 2024, MNRAS, 527, 4730; 7) Diaz-Luis et al., 2019, A&A, 629, A94; 8) Mac Leod et al., 2023, ApJ, 956, 27; 9) Famaey et al., 2009, A&A, 498, 627; 10) McDonald et al., 2012, MNRAS, 427, 343; 11) Baek et al., 2025, AJ, 170, 84; 12) Skinner et al., 1990, MNRAS, 243, 78; 13) Winters et al., 2022, A&A, 658, 135; 14) Libert et al., 2010, A&A, 515A, 112L; 15) Neilson et al., 2014, A&A, 568, 88; 16) Buchler et al., 2004, ApJ, 613, 532; 17) Power et al., 2025, arXiv:2509.11672; 18) Neilson et al., 2023, A&A, 677, A96 ; 19) Kiss et al., 2000, AASS, 145, 283; 20) Hartig et al., 2014, AJ, 148, 123; 21) Ahmad et al., 2025, A&A, 699, 148; 22) Banyai et al., 2013, MNRAS, 436, 1576; 23) Shetye et al., 2025, A&A, 702, A39; 24) Lebzelter & Hron, 2003, A&A, 411, 533; 25) Lebzelter et al., 2019, ApJ, 886, 117; 26) Takeuti et al., 2013, PASJ, 65, 60; 27) Lebzelter & Hron, 1999, AA, 351, 533; 28) Andronov & Chinarova, 2012, Odessa Astron. Pub., 25, 148; 29) Kiss et al., 1999, AA, 346, 542; 30) Danilovich et al., 2025, A&A, 704, A341; 31) Montargès et al., 2023, A&A, 671, A96; 32) Wallström et al., 2024, A&A, 681, 50; 33) Baudry et al., 2023, A&A, 674, 125; 34) Soszyński et al., 2021, ApJ, 911, 22; 35) Goldberg et al., 2024, ApJ, 907, 35; 36) Glass & Van Leeuwen, 2007, MNRAS, 378, 1543; 37) Tsuji, 2008, A&A, 489, 1271; 38) Halabi & El Eid, 2015, MNRAS, 451, 2957; 39) Olofsson et al., 2002, A&A, 391, 1053; 40) Van Belle et al., 2013, ApJ, 775, 45; 41) Hedrosa et al., 2013, ApJL, 768, L11; 42) Matthews et al., 2011, ApJ, 141, 60; 43) Jorissen et al., 2011, A&A, 532, 135 ; 44) Castro-Carrizo et al., 2010, A&A, 523, 59; 45) De Vicente et al., 2016, A&A, 589, 7; 46) Nakashima, 2019, ApJ, 620, 943; 47) Dumm & Schild, 1998, New Astron., 3, 137; 48) Lebzelter & Hilke, 2002, A&A, 393, 563; 49) Ghosh et al., 2019, MNRAS, 484, 4619; 50) Gottlieb et al., 2022, A&A, 660, 94; 51) Baier et al., 2010, A&A, 516, 45; 52) Chen et al., 2016, MNRAS, 460, 4182; 53) Hoai et al., 2022, VJSTE, 64, 16; 54) Kervella et al., 2016, A&A, 596, A92; 55) Binder et al., 2008, ApJ, 685, L145; 56) McDonald et al., 2020, MNRAS, 491, 1174; 57) Templeton et al., 2008, JAAVSO, 36/1, 1; 58) Uttenthaler et al., 2011, A&A, 531, A88.

**Table A1.10** Family Byes

| Name | Type | *C* | *A* | *T* | *L* | *R* | $\dot{M}$ | Tc | $^{12}C/^{13}C$ | Ref | *P* | *W* | *q* |
|---|---|---|---|---|---|---|---|---|---|---|---|---|---|
| 158 GY Aql | M8 | 2.7 | 4.1 | 3100 | | | 23 | - | | 1,2,3,4,5 | 470 | - | - |
| 159 V Boo | M5 | 1.3 | 2.3 | | | | | - | | 6,7 | 257 | 0.49 | -0.034 |
| 160 R Cam | S2 | 0.4 | 4.5 | | | | | - | | 8 | 270 | 0.41 | -0.139 |
| 161 T Cam | S5 | 0.7 | 5.4 | | | | | yes | 31 | - | 376 | 0.40 | -0.169 |
| 162 T Cas | M7 | 1.9 | 4.3 | | | | | yes | 33 | 6,9,10 | 445 | 0.39 | -0.095 |
| 163 T Cep | M6 | 1.2 | 4.1 | 2400 | 5.7 | | 1 | yes | 33 | 11 | 338 | 0.33 | -0.049 |
| 164 S CMi | M7 | 1.9 | 4.9 | 2800 | | | 0.5 | yes | 18 | 11 | 333 | 0.32 | -0.060 |
| 165 V Cnc | S0 | 1.4 | 5.0 | 2060 | | | | yes | | - | 272 | 0.33 | -0.069 |
| 166 FF Cyg | S5.5 | 1.0 | 4.5 | | | | | - | | - | 328 | 0.34 | -0.093 |

| 167 RZ Cyg | M7 | 2.6 | 2.5 | 3050 | | | | - | 40 | 12 | 280 | - | - |
|---|---|---|---|---|---|---|---|---|---|---|---|---|---|
| 168 Z Del | S4.5 | 1.4 | 5.5 | 2800 | | | | yes | | - | 305 | 0.30 | -0.012 |
| 169 R Gem | S3.5 | 2.0 | 6.2 | 2400 | 5.5 | | 4 | yes | 22 | 11 | 361 | 0.31 | -0.035 |
| 170 T Gem | S2 | 0.5 | 3.4 | 3150 | | | | yes | | - | 287 | 0.38 | -0.139 |
| 171 S Her | M4 | 1.1 | 5.4 | 3100 | | | | yes | | 13 | 307 | 0.31 | -0.077 |
| 172 R Hya | M6 | 0.9 | 3.7 | 2500 | | | 4 | yes | 26 | 4,14 | 386 | 0.36 | -0.035 |
| 173 W Hya | M7.5 | 0.4 | 3.2 | 2600 | 7.0 | 430 | 0.8 | no | 16 | 15,16,17,18 | 390 | 0.37 | 0.009 |
| 174 R Lyn | S4 | 1.4 | 5.7 | 2400 | 5.6 | | 3 | yes | | 19,20 | 378 | 0.26 | -0.036 |
| 175 X Oct | M5/6 | 1.9 | 3.1 | | | | | - | | - | 201 | 0.46 | -0.086 |
| 176 Z Peg | M7 | 1.7 | 4.7 | | | | | yes | | - | 326 | 0.30 | -0.014 |
| 177 U Per | M5.5 | 1.6 | 3.2 | 3100 | 9.0 | | | yes | | 6,8,21,22,23 | 321 | 0.47 | -0.104 |
| 178 S UMa | S0.5 | 1.2 | 3.7 | | | | | yes | | 24 | 226 | 0.46 | -0.147 |
| 179 U UMi | M6 | 1.6 | 3.3 | | | | | - | | - | 324 | 0.37 | -0.034 |

1) Montargès et al., 2023, A&A, 671, A96; 2) Wallström et al., 2024, A&A, 681, 50; 3) Gottlieb et al., 2022, A&A, 660, 94; 4) Baudry et al., 2023, A&A, 674, 125; 5) Loup et al., 1993, A&AS, 99, 291L; 6) Shetye et al., 2025, A&A, 702, A39; 7) Lebzelter & Hron, 1999, AA, 351, 533; 8) Hoai et al., 2025, ApJ, 982, 201; 9) Ramstedt et al., 2012, A&A, 543, A147; 10) Merchán Benitez et al., 2023, A&A, 672, 165; 11) Danilovich et al., 2015, A&A, 581, A60; 12) Hinkle et al., 2016, ApJ, 825, 38H; 13) Uttenthaler et al., 2011, A&A, 531, A88; 14) Fadeyev, 2023, INASR, 8, 69; 15) Ohnaka et al., 2025, A&A, 704, A18; 16) Lebzelter et al., 2025, A&A, 431, 623; 17) Hoai et al., 2022, VJSTE, 64, 16; 18) Glass & Van Leeuwen, 2007, MNRAS, 378, 1543; 19) Uttenthaler et al., 2024, A&A, 690, 393; 20) Ramstedt et al., 2009, A&A, 499, 515; 21) Nhung et al. 2025, ApJ, 994, 130; 22) Cunha et al., 2020, MNRAS, 499, 4687; 23) Cadmus, 2021, Astron. J., 161, 75; 24) Famaey et al., 2009, A&A, 498, 627; 22) Zacs 2015;

**Table A1.11** Family C1

| Name | Type | *C* | *A* | *T* | *L* | *R* | $\dot{M}$ | Tc | $^{12}C/^{13}C$ | Ref | *P* | *W* | *q* |
|---|---|---|---|---|---|---|---|---|---|---|---|---|---|
| 180 AQ And | CN5 | 1.4 | 0.4 | 2740 | | | 4 | | 30 | 1,2,3,4 | 330 | | |
| 181 VX And | CJ4.5 | 1.5 | 0.9 | 2520 | | | 1 | | 13 | 3,5,6 | 370 | | |
| 182 V Aql | CN5 | 1.4 | 0.3 | 2750 | 6.5 | | 4 | | 82 | 3,7,8,9 | | | |
| 183 UU Aur | CN5 | 1.5 | 0.4 | 2700 | 6.9 | | 4 | | 80 | 3,7,10,11,12,13,14 | | | |
| 184 ST Cam | CN5 | 1.5 | 0.3 | 2800 | | | 11 | | 61 | 3,10,15,16 | | | |
| 185 U Cam | C3 | 1.8 | 0.4 | 2750 | 7.0 | 290 | | | 97 | 3,10,17,18,19,20,21 | 220 | | |
| 186 UV Cam | CJ4 | 0.9 | 0.3 | 3200 | | | | | 4 | 22 | | | |
| 187 RT Cap | C6.4 | 1.0 | 0.7 | 2480 | 2.2 | | 2 | | 59 | 3,10,23,24,25 | | | |
| 188 T Cnc | CN5 | 1.3 | 0.7 | 2380 | | | 10 | | | 3,26,27 | | | |
| 189 X Cnc | CN4.5 | 1.1 | 0.3 | 2700 | | | 4 | | 50 | 3,10,28,29,30 | | | |
| 190 Y CVn | CN5/J | 1.4 | 0.3 | 2730 | 7.0 | | 2 | | 3 | 3,13,31,32,33,25 | | | |
| 191 AW Cyg | C4,5 | 1.6 | 0.3 | 2760 | | | | | 21 | 1,2 | | | |
| 192 RV Cyg | CN5 | 1.9 | 0.3 | 2675 | 13.4 | | 8 | | 74 | 3,10,34 | | | |
| 193 TT Cyg | C5,4 | 1.1 | 0.3 | 2900 | 2.7 | | 1 | | | 3,4,15,19,35,36 | | | |
| 194 V460 Cyg | CN5 | 1.2 | 0.2 | 2950 | | | 5 | | 40 | 3,10,37,38 | | | |
| 195 RY Dra | C4J | 1.5 | 0.4 | 2300 | 4.5 | | 2 | | 3 | 3,10,14,39 | | | |
| 196 UX Dra | C | 1.1 | 0.4 | 3090 | | | 4 | | 28 | 3,5,40 | | | |
| 197 SY Eri | CN5 | 1.1 | 0.3 | | | | | | | - | | | |
| 198 TU Gem | CN5 | 1.5 | 0.3 | 2950 | | | 4 | | 59 | 3,10,21,41 | | | |
| 199 U Hya | CN5 | 1.5 | 0.3 | 2800 | 3.8 | | 3 | yes | 32 | 3,710,42,43,44,45,46 | | | |
| 200 RV Mon | C4,5 | 1.7 | 0.3 | 3050 | | | 2 | | | 3,47 | | | |
| 201 RT Ori | CN5 | 1.0 | 0.3 | 2870 | | | 2 | | | 3 | | | |
| 202 V431 Ori | C5,5 | 1.6 | 0.4 | 2540 | | | 2 | | | 3 | | | |
| 203 W Ori | CN5 | 1.3 | 0.5 | 2650 | 6.8 | | 3 | no | 44 | 3,8,10,11,34,42,48,499,50 | | | |
| 204 TX Psc | CN6 | 1.0 | 0.2 | 3100 | 7.9 | | 6 | yes | 28 | 3,10,48,51,52,53 | | | |
| 205 Z Psc | CN5 | 1.0 | 0.3 | 3100 | | | 1 | | 55 | 3,10,47 | | | |
| 206 S Sct | CN5 | 1.1 | 0.3 | 2600 | 4.5 | | | | 45 | 3,10,23,51,54,55,56 | | | |
| 207 Y Tau | CN5 | 2.2 | 0.4 | 2700 | | | 2 | | 58 | 3,10,16,57,58 | | | |

1) Van Belle et al., 2013, ApJ, 775, 45; 2) Hedrosa et al., 2013, ApJL, 768, L11 ; 3) Bergeat & Chevallier, 2005, A&A, 429, 235; 4) Kerschbaum et al., 2010, A&A, 518, 140; 5) Hinkle et al., 2016, ApJ, 825, 38H; 6) Choplin et al., 2024, A&A, 691, 7; 7) Danilovich et al., 2015, A&A, 581, A60; 8) Jeste et al., 2022, A&A, 666, A69; 9) Smirnova, 2012, MNRAS, 424, 2468; 10) Abia et al., 2017, A&A, 599, A39; 11) Schöier & Olofsson, 2001, A&A, 368, 969; 12) Schöier et al., 2013, A&A, 550, A78; 13) Milam et al., 2009, ApJ, 690, 837; 14) Ramstedt & Olofsson, 2014, A&A, 566, 145; 15) Kraemer et al., 2019, ApJ, 887, 82; 16) Lopez & Hiriart, 2011, AJ, 142, 11; 17) Olofsson et al., 2010, A&A, 515, A27; 18) Ramstedt et al., 2011, A&A, 531, 148; 19) Cox et al., 2012, A&A, 537, 35; 20) Massalkhi et al., 2018, A&A, 611, 29; 21) Tanaka et al., 2007, PASJ, 59, 939; 22) Zacs et al., 2015, ApJ, 803, 17; 23) Mecina et al., 2014, A&A, 566, A69; 24) Baug & Chandrasekhar, 2012, MNRAS, 419, 866; 25) Kastner & Wilson, 2021, ApJ, 922, 24; 26) Richichi et al., 1991, AA, 241, 131; 27) Bogdanov, 1994, Astron. Lett., 20, 221 ; 28) Famaey et al., 2009, A&A, 498, 627; 29) Menten et al., 2018, A&A, 613, A49; 30) Ramstedt & Olofsson, 2014, A&A, 566, 145; 31) Matthews et al., 2013, AJ, 145, 97; 32) Libert et al., 2007, MNRAS, 380, 1161; 33) Fonfria et al., 2021, A&A, 651, A8; 34) Lombaert et al., 2016, A&A, 588, A124; 35) Maercker et al., 2022, A&A, 663, A64; 36) Groenewegen, 2012, A&A, 543, 36; 37) Dominy et al., 1978, ApJ, 223, 949; 38) Little-Marenin & Little, 1984, ApJ, 283, 188; 39) Kwok et al., 1997, ApJSS, 112, 557; 40) Mecina et al., 2012, ASP Conf. Series, Vol. 445; 41) Richichi & Chandrasekhar, 2006, A&A, 451, 1041; 42) Mecina et al., 2020, A&A, 644, A66; 43) Izumiura et al., 2011, A&A, 528, 29; 44) Rau et al., 2017, A&A, 600, 92; 45) Dharmawardena et al., 2018, MNRAS, 479, 536; 46) Danilovich et al., 2018, A&A, 617, A132; 47) Paladini et al., 2011, A&A, 533, A27; 48) Cruzalèbes et al., 2013, MNRAS, 434, 437; 49) Cruzalèbes et al., 2015, MNRAS, 446, 3277; 50) Lebzelter & Hron, 2003, A&A, 411, 533; 51) Paladini et al., 2017, A&A, 600, A136 ; 52) Brunner et al., 2019, A&A, 621, A50; 53) Klotz et al., 2013, A&A, 550, A86; 54) Maercker et al., 2024, A&A, 687, A112; 55) Bergman & Lindqvist, 2015, A&A, 582, A102; 56) De Beck et al., 2010, A&A, 523, 18; 57) Richichi et al., 1995, A7A, 301, 439; 58) Dominy & Wallerstein, 1986, PASP, 98, 1193.

**Table A1.12** Family C2

| Name | Type | *C* | *A* | *T* | *L* | *R* | $\dot{M}$ | Tc | $^{12}C/^{13}C$ | Ref | *P* | *W* | *q* |
|---|---|---|---|---|---|---|---|---|---|---|---|---|---|
| 208 ST And | C5.4 | 0.6 | 1.8 | | | | 2 | | | 1 | 340 | | |
| 209 UV Aur | C8 | 2.8 | 2.6 | | | | | yes | | 2 | 395 | 0.43 | |
| 210 V Aur | C6 | 1.6 | 2.6 | 2820 | | | 36 | | | - | 353 | 0.45 | |
| 211 S Cam | C6 | 1.1 | 1.9 | | | | | | 14 | 3 | 330 | 0.56 | -0.146 |
| 212 W Cas | C9 | 1.0 | 2.8 | | | | | yes | 25 | - | 408 | 0.46 | -0.079 |
| 213 RV Cen | C3 | 1.6 | 2.4 | 2860 | | | 8 | | | - | 445 | 0.51 | -0.099 |
| 214 PQ Cep | C6 | 2.2 | 2.6 | | | | | | | - | 443 | 0.46 | -0.041 |
| 215 R CMi | C0 | 1.3 | 2.9 | 3000 | | | | yes | 19 | 4 | 340 | 0.47 | |
| 216 BH Cru | C8 | 1.5 | 2.9 | 3000 | | | | yes | 8 | 5,6,7,8,9,10 | 513 | 0.47 | -0.049 |
| 217 RS Cyg | C5.5 | 1.2 | 1.6 | 3100 | | | 2 | | | 11,12,13 | 419 | 0.66 | -0.127 |
| 218 U Cyg | C9 | 2.1 | 3.1 | 2650 | | | 10 | yes | 14 | 14,15 | 463 | 0.39 | -0.019 |
| 219 WX Cyg | C6 | 1.5 | 2.6 | | | | | yes | 5 | 16 | 411 | 0.52 | |
| 220 T Lyn | C7 | 2.0 | 2.8 | | | | 4 | | | 1; 9 | 409 | 0.44 | |
| 221 U Lyr | C4 | 2.1 | 1.8 | | | | | | 23 | - | 452 | 0.54 | |
| 222 V Oph | C4 | 1.5 | 2.4 | 2800 | 3.6 | 470 | 10 | yes | 11 | 17,18,19,20 | 297 | 0.52 | |
| 223 R Ori | SC5 | 1.8 | 3.5 | 3080 | | | 6 | | | - | 377 | 0.32 | 0.011 |
| 224 SY Per | C6.4 | 1.9 | 2.0 | 2705 | | | 15 | | 43 | 12,21,22 | 480 | | |
| 225 Y Per | C5 | 1.4 | 1.9 | 3150 | | | 3 | | | - | 249 | 0.55 | |
| 226 RZ Peg | SC5 | 2.3 | 4.1 | 2925 | | | 1 | yes | 9 | 20 | 437 | 0.33 | -0.012 |
| 227 R Scl | CN5 | 2.1 | 1.5 | 2640 | 6.3 | | 12 | | | 21,23,24,25,26,27, 28,29,30,31,32,33 | 373 | | |
| 228 SS Vir | CN4.5 | 2.1 | 1.8 | 2500 | | | 4 | | | 21,34 | 361 | 0.44 | -0.010 |
| 229 BD Vul | C6 | 1.7 | 2.3 | | | | | | | - | 430 | 0.45 | |

1) Bogdanov & Taranova, 2012, Astron. Rep., 56, 124; 2) Luna et al., 2013, A&A, 559, 6; 3) Tanaka et al., 2007, PASJ, 59, 939; 4) Abia et al., 2019, A&A, 625, 40; 5) Uttenthaler et al., 2011, A&A, 531, A88; 6) Uttenthaler et al., 2016, A&A, 585, 145; 7) Uttenthaler et al., 2019, A&A, 622, A120; 8) Uttenthaler & Merchan-Benitez, 2025, A&A, 700, 187; 9) Merchán Benitez et al., 2023, A&A, 672, 165; 10) Zijlstra et al., 2004, MNRAS, 352, 325; 11) Binder et al., 2008, ApJ, 685, L145; 12) Reiter et al., 2015, MNRAS, 447, 3909; 13) Cadmus, 2022, AJ, 163, 265; 14) Lopez & Hiriart, 2011, AJ, 142, 11; 15) Hinkle, 2016, ApJ, 825, 38H; 16) Whitelock et al., 2008, MNRAS, 386, 313; 17) Rau et al., 2019, ApJ, 882, 37; 18) Ohnaka et al., 2007, A&A, 466, 1099; 19) Hulberg et al., 2025, ApJ, 982,135; 20) Guandalini & Cristallo, 2013, A&A, 555, A120; 21) Bergeat & Chevallier, 2005, A&A, 429, 235; 22) Hedrosa et al., 2013, ApJL, 768, L11; 23) Wittkowski et al., 2017, A&A, 601, A3; 24) Sacuto et al., 2011, A&A, 525, A42; 25) Hankins et al., 2018, ApJ, 852, 27H; 26) Olofsson et al., 2010, A&A, 515, A27; 27) Maercker et al., 2014, A&A, 570, 101; 28) Saberi et al., 2017, A&A, 599, 63; 29) Abia et al., 2017, A&A, 599, A39; 30) Maercker et al., 2016, A&A, 586, 5; 31) Drevon et al., 2022, A&A, 665, 32; 32) Cruzalèbes et al., 2013, MNRAS, 434, 437; 33) Cruzalèbes et al., 2015, MNRAS, 446, 3277; 34) Richichi & Chandrasekhar, 2006, A&A, 451, 1041.

**Table A1.13** Family C3

| Name | Type | *C* | *A* | *T* | *L* | *R* | $\dot{M}$ | Tc | $^{12}C/^{13}C$ | Ref | *P* | *W* | *q* |
|---|---|---|---|---|---|---|---|---|---|---|---|---|---|
| 230 AZ Aur | C8 | 2.5 | 3.3 | | | | | yes | | - | 415 | 0.37 | |
| 231 S Aur | C4 | 3.4 | 2.0 | 1940 | 17 | | 45 | | | 1,2,3,4 | 610 | | |
| 232 R Cap | C | 3.0 | 3.5 | | | | | | | - | 342 | 0.35 | 0.038 |
| 233 X Cas | C5 | 1.5 | 1.9 | | | | | | 19 | - | 424 | 0.52 | |
| 234 AX Cep | C | 2.7 | 2.9 | | | | | | | - | 396 | 0.39 | 0.026 |
| 235 S Cep | C7 | 2.2 | 2.4 | 2240 | 7.3 | | 20 | yes | 224 | 5,6 | 484 | 0.40 | 0.033 |
| 236 V CrB | C6 | 2.1 | 3.0 | 2090 | 5.3 | | 8 | yes | | 7,8 | 358 | 0.37 | 0.033 |
| 237 LX Cyg | C | 1.7 | 3.4 | | | | | yes | 32 | 9,10 | 581 | 0.46 | 0.042 |
| 238 V Cyg | C7 | 2.8 | 3.4 | 2200 | 8.5 | 480 | 37 | | 20 | 5,11,12 | 420 | 0.38 | 0.029 |
| 239 T Dra | C6 | 3.0 | 3.0 | 1850 | | | 53 | yes | 24 | 13,14 | 422 | 0.37 | 0.046 |
| 240 R For | C4 | 2.7 | 3.0 | 2000 | 5.8 | | 24 | | | 15,16,17 | 387 | 0.35 | 0.052 |
| 241 VX Gem | C9 | 1.6 | 4.0 | | | | | yes | 9 | - | 379 | 0.36 | 0.035 |
| 242 V Hya/K | C6 | 2.7 | 1.5 | 2400 | 14 | | 50 | | 70 | 3,18 | 510 | | |
| 243 R Lep | C7 | 2.5 | 2.5 | 2300 | 6.0 | | 15 | | 34 | 8,19,20,21,22,23 | 440 | 0.42 | 0.012 |
| 244 RU Vir | C8 | 3.7 | 3.0 | 2100 | | | 23 | yes | 19 | 24,25,26 | 437 | 0.37 | 0.043 |
| 245 R Vol | C | 3.4 | 2.9 | | 8.3 | | 22 | | | 2,20,27 | 451 | 0.31 | 0.087 |

1) Lombaert et al., 2016, A&A, 588, A124; 2) Jeste et al., 2022, A&A, 666, A69; 3) Bergeat & Chevallier, 2005, A&A, 429, 235; 4) Lopez & Hiriart, 2011, AJ, 142, 11; 5) Castro-Carrizo et al., 2010, A&A, 523, 59; 6) Nowotny et al., 2010, A&A, 514, A35; 7) Libert et al., 2010, A&A, 515A, 112L; 8) Danilovich et al., 2015, A&A, 581, A60; 9) Uttenthaler et al., 2016, A&A, 585, 145; 10) Zijlstra et al., 2004, MNRAS, 352, 325; 11) Neufeld et al., 2010, A&A, 521, L5; 12) Decin et al., 2008, A&A, 480, 431D; 13) Alonso-Hernández et al., 2024, A&A, 684, A77; 14) Alonso-Hernández et al., 2025, A&A, 698, A319; 15) Guerrero & Ortiz, 2024, MNRAS, 527, 4730; 16) Schöier et al., 2013, A&A, 550, A78; 17) Lobel et al., 1999, AA, 343, 46; 18) Milam et al., 2009, ApJ, 690, 837; 19) Paladini et al., 2017, A&A, 600, A136; 20) Rau et al., 2017, A&A, 600, 92; 21) Danilovich et al., 2018, A&A, 617, A132; 22) Hinkle et al., 2016, ApJ, 825, 38H; 23) Ortiz et al., 2022, Proc. IAU Symp. 366; 24) Rau et al., 2015, A&A, 583A, 106R; 25) Nowotny et al., 2011, A&A, 529, 12; 26) Nowotny et al., 2013, A&A, 552, 20; 27) Ramstedt et al., 2020, A&A, 640, 133.